\documentclass[svgnames]{aa}
\bibpunct{(}{)}{,}{a}{}{,}  

\newcommand{\Msol}{\ensuremath{\,{\rm M}_{\sun}}}

\usepackage{graphicx}
\usepackage{txfonts}
\usepackage[svgnames]{xcolor}
\graphicspath{{./}{figures/}}

\begin{document}

\title{A new $^{12}$C~+~$^{12}$C nuclear reaction rate: impact on stellar evolution}

\author{E. Monpribat\inst{1} 
        \and S. Martinet\inst{2}
        \and S. Courtin \inst{1,3}
        \and M. Heine\inst{1}
        \and S. Ekstr\"om\inst{2}
        \and D. G. Jenkins\inst{3,4}
        \and A. Choplin\inst{5}
        \and P. Adsley\inst{6,7}
        \and D. Curien\inst{1}
        \and M. Moukaddam\inst{1}
        \and J. Nippert\inst{1}
        \and S. Tsiatsiou\inst{2}
        \and G. Meynet\inst{2}
        }
\institute{Université de Strasbourg, CNRS, IPHC UMR 7178, F-67000 Strasbourg, France \\ \email{sandrine.courtin@iphc.cnrs.fr} 
\and Department of astronomy, University of Geneva, chemin Pegasi 51, 1290 Versoix, Switzerland \\
\email{sebastien.martinet@unige.ch} 
\and University of Strasbourg Institute of Advanced Studies (USIAS), Strasbourg, France
\and Department of Physics, University of York, Heslington, York YO10 5DD, United Kingdom
\and Institut d'Astronomie et d'Astrophysique, Universit\'e Libre de Bruxelles, CP 226, 1050, Brussels, Belgium
\and Cyclotron Institute, Texas A\&M University, College Station, Texas 77843, USA
\and Department of Physics \& Astronomy, Texas A\&M University, College Station, Texas 77843, USA}

 \abstract
    {By changing the internal composition of stars, nuclear reactions play a key role in their evolution and in their contribution to the chemical evolution of galaxies. Recently,
    the STELLA collaboration has carried out new direct measurements
    of the $^{12}$C~+~$^{12}$C fusion cross section, one of the key reactions occurring in C-burning regions in massive stars. Using a coincidence technique, accurate measurements were obtained for many different energies with the lowest energy explored corresponding to the Gamow window for massive stars.}
    %    Stellar nucleosynthesis is one of the most important mechanisms in the formation and evolution of the Universe as it is known today. 
    %    Of the nuclear reactions taking place in stars, some are more important than others for stellar evolution: an example being the $^{12}\mathrm{C}$ + $^{12}\mathrm{C}$ fusion reaction. 
    %    The later, difficult to study experimentally at the astrophysical region of interest, is currently at the center of research of the STELLA collaboration.}
    {This work presents new  $^{12}$C~+~$^{12}$C reaction rates in the form of numerical tables with associated uncertainty estimation, as well as analytical formulae that can be directly implemented into stellar evolution codes. This article further describes the impact of these new rates on C-burning in stars.}
%    aim to determine new reaction rates relevant for astrophysical interest, based on direct measurements of the fusion cross sections from the STELLA experiment, and investigate the impact of these reaction rates on stars evolution.
    {We determine reaction rates for two cross-section extrapolation models: one based on the fusion-hindrance phenomenon, and the other on fusion-hindrance plus a resonance, and compare our results to previous data. Using the GENEC stellar evolution code, we study how these new rates impact the C-burning phases in two sets of stellar models for stars with 12\Msol\ and 25\Msol\ initial masses chosen to be highly representative of the diversity of massive stars.} 
  % results heading (mandatory) sry..done
    {The effective temperatures of C-burning in both sets of stellar models are entirely covered by the sensitivity of the present experimental data, and no extrapolation of the rates is required.
    %The effective temperatures of C-burning in both sets of stellar models are almost entirely covered by the sensitivity of the present experimental data and only modest extrapolation of the rates is required.
    Although, the rates may differ by more than an order of magnitude for temperatures typical of C-burning, the impacts on the stellar structures during that phase remain modest. This is a consequence of the readjustment of the stellar structure to a change of nuclear reaction rate for reactions important for energy production. 
%    Indeed, the star readjusts its internal conditions until similar energy is produced as before the rate has been changed. In general these readjustments are very small due to the high dependence of the nuclear reaction rates on the temperature. 
For the hindrance case, the C-burning phase is found to occur at central temperatures 10\% higher than with the hindrance plus resonance rate. Its C-burning lifetime is reduced by a factor of two. This model, nevertheless, loses more entropy than the other one thus enters earlier into the degeneracy regime which will impact the last stages of the evolution at the pre-core collapse time. The hindrance model produces up to 60\% more neon. The impact of the different rates on the s-process occurring during the C-burning phase is modest, changing final abundances of s-processed elements by at most 20\% (cobalt).}
%    We note however that the models using hindrance rates have  carbon-burning lifetimes reduced by a factor 2, which may actually have important consequences for the end phases of the evolution. Indeed shorter carbon-burning phases imply less entropy lost by the core through neutrino emissions during that phase which may impacted the final fate of the star.}
    {}

\maketitle

\section{Introduction} \label{sec:intro}

The origin of the chemical elements in our Universe is one of the key questions driving contemporary subatomic physics. The Big Bang produces only hydrogen, helium and trace amounts of lithium. The remaining 80 chemical elements in the Mendeleev Table are believed to be forged through nuclear reactions in stars. In many cases, however, our understanding of the relevant nuclear reactions is incomplete; examples include the fusion processes taking place in a star like the sun \citep{Villante2021}, and the fusion of oxygen in massive stars \citep{Holt2019}. 
%In particular, the abundances of carbon and oxygen, deeply connected with the late stages of stellar evolution --- quiescent or explosive --- are constrained by the probability for nuclear fusion at stellar energies. 
In this paper, we focus on carbon-fusion reactions that are the main energy source in the C-burning regions in a star. 
Let us recall that the evolution of stars starts with H-burning, followed by He-burning, during which carbon is synthesized, through the triple-alpha process. If the star is massive enough (with initial masses higher than $\sim$8\Msol), its temperature and density further increase and C-burning may start (typically at temperatures above 0.6-0.8 GK), first at the centre of the star, and then in a shell. 

The carbon fusion reaction, $^{12}$C + $^{12}$C, leads to the formation of a $^{24}$Mg$^*$ compound nucleus at an excitation energy around 15-17~MeV; 
%The decay $^{24}$Mg$^*$ can occur along different channels, the most important ones being
subsequent particle emission leads to the residual nuclei, $^{20}$Ne, $^{23}$Na and $^{23}$Mg, through the following set of processes:
%The carbon fusion reactions produce different elements depending on the
%way the compound nucleus $^{24}$Mg$^*$ decays:
\begin{align}
^{12}\mathrm{C} + ^{12}\mathrm{C} \, \rightarrow \, ^{24}\mathrm{Mg}^{*} \, & \rightarrow \, ^{20}\mathrm{Ne} + \alpha && (Q=4.62 \, \mathrm{MeV}), \nonumber\\
                                                                            & \rightarrow \, ^{23}\mathrm{Na} + p  && (Q=2.24 \, \mathrm{MeV}), \nonumber \\
                                                                            & \rightarrow \, ^{23}\mathrm{Mg} + n  && (Q=-2.62 \, \mathrm{MeV}).
%                                                                            & \rightarrow \, ^{16}\mathrm{O} + ^{16}\mathrm{O}  && (Q=-0.11 \, \mathrm{MeV}). 
%                                                                            & \rightarrow \, ^{24}\mathrm{Mg} + \gamma && (Q=13.93 \, \mathrm{MeV}), \nonumber
\end{align}
In an astrophysical context, the dominant exit channels are the alpha and proton channels, the neutron emission channel being endothermic \citep{Bucher2015}. 
Together with these main fusion reactions, other reactions are important during the C-burning phase \citep[see e.g. the book by][]{Illiadis}. An important one is the reaction $^{23}$Na(p, $\alpha$)$^{20}$Ne that converts nearly all the $^{23}$Na into $^{20}$Ne. The elements most strongly produced by the C-burning phase are $^{20}$Ne and $^{24}$Mg.

The $^{12}$C~+~$^{12}$C fusion reaction has been extensively studied, for more than fifty years, using direct measurements \citep[see e.g.][]{Patterson1969,High1977,Becker1981,Aguilera2006,Barron2006,Spillane2007,Bucher2015}. Since the nuclei produced by the decay of $^{24}$Mg$^*$, $^{20}$Ne, $^{23}$Na and $^{23}$Mg nuclei are often produced in excited states, and then emit gamma rays, measurements of carbon fusion involves detection of light particles and/or de-excitation gamma rays.
The studies cited above were all based on the detection of either light charged particles or gamma rays. The results obtained were consistent for center of mass energies above and near the Coulomb barrier. However, for energies below the Coulomb barrier ---  the astrophysically interesting region where cross sections are small, down to the picobarn range --- the associated astrophysical S-factors from different measurements often differ by several orders of magnitude. The uncertainties inherent in the published data make it difficult to perform model extrapolations down to the lowest energies \citep{Li2020}. 
Difficulties in extrapolating the S-factors mainly come from two different sources. First, unlike most heavy-ion fusion reactions, the $^{12}$C~+~$^{12}$C fusion cross section exhibits strongly oscillatory or resonant behavior on and below the Coulomb barrier \citep{Almqvist1960}. The narrow resonances in the cross section have traditionally been attributed to $^{24}$Mg excited states with a special $^{12}$C-$^{12}$C molecular configuration \citep{Jenkins2015}, while, more recently, it has been speculated that the origin of the oscillations is due to the low density of states in the $^{24}$Mg compound nucleus \citep{Jiang2013}.
The second difficulty comes from experimental backgrounds which are, from a practical point-of-view, unavoidable in such low cross section studies at the energies relevant to massive stars (see sect. \ref{subsec:se}). 

In the last decade, to reduce such background, \citet{Jiang2012} proposed a coincidence detection method, tailored to studies of $^{12}$C~+~$^{12}$C fusion; first experimental results have been published in \citet{Jiang2018}. More recently, this coincidence method has been used in two different direct measurements by \citet{Fruet2020} and \citet{Tan2020}, and show consistent results. In this paper, we present new accurate reaction rates that can be incorporated into astrophysical models, based on direct measurements performed with a coincidence method, that have been obtained for centre of mass energy in the Gamow window for massive stars. The cross section data employed derives from the direct measurement recently published by \citet{Fruet2020} which was obtained with the STELLA experiment described more fully in Sect.~\ref{subsec:se}.
%At the same time, indirect measurement techniques using the Trojan Horse Method (THM) have been developed, and new cross sections for the $^{12}$C~+~$^{12}$C fusion reaction have been reported and discussed \citep{Tumino2018,Mukhamedzhanov2019}.

New stellar models for representative masses in the massive star range
are then computed with these new rates during the C-burning phase. Carbon-burning in stars has been studied since a long time \citep[see e.g.][]{Arnett1969, Lamb1976, Weaver1978, Sparks1980, Arnett1985} but, interestingly, it still remains a topic of lively research today. For instance, \citet{Gasques2007, Pignatari2013} discussed the impact of the $^{12}$C~+~$^{12}$C rates on the s-process occurring in C-burning regions; \citet{Sukhbold2020} studied the impact of the core C-burning phase on the final fate of red supergiants; \citet{Fields2018}, in a very exhaustive study of the impact of nuclear uncertainties in models for pre-supernovae, identified, among the reactions having the largest impact on the pre-supernova structure, the reaction $^{12}$C($^{12}$C,p)$^{23}$Na; \citet{Chieffi2021} have investigated the impact of new rates for the carbon fusion reaction in stellar models.
%These recent works illustrates that carbon burning is thus a lively research line.

%through the following set of processes:
%
%\begin{align}
%^{12}\mathrm{C} + ^{12}\mathrm{C} \, \rightarrow \, ^{24}\mathrm{Mg}^{*} \, & \rightarrow \, ^{20}\mathrm{Ne} + \alpha && (Q=4.62 \, \mathrm{MeV}), \nonumber \\
%& \rightarrow \, ^{23}\mathrm{Na} + p  && (Q=2.24 \, \mathrm{MeV}), \nonumber \\
%& \rightarrow \, ^{23}\mathrm{Mg} + n  && (Q=-2.62 \, \mathrm{MeV}). 
%\end{align}
%

As said above, in the present work, our aim is to explore the consequences of new rates
for the carbon fusion reaction in stellar models. We shall also use 
simpler models, where we focus on the detailed abundances variations resulting from a very large nuclear reaction network (following more than 1400 isotopes). 
Compared to the recent work by \citet{Chieffi2021}, we use here different experimental data for the nuclear reaction rates that, as we shall see, lead to very different consequences.

The paper begins by briefly presenting the nuclear experiment in Sect.~2.
Section 3 discusses the results, and the tables and formulae that can be directly implemented in stellar evolution codes. Section 4 and 5 discuss the impact of the new rates in, respectively, stellar models and in a simpler model (``one-layer model'') with a very extended nuclear reaction network. Finally, a brief summary with some perspectives for future work are given in Sect.~6.

\section{Direct measurement of fusion reactions of astrophysics interest: the STELLA experiment}

\subsection{A brief historical overview}

Fusion is among the most studied processes in the collision of heavy ions at nuclear physics energies. The interactions between two colliding nuclei are, in general, described using the Coulomb and nuclear potentials, the sum of these forming the so-called Coulomb barrier (CB) which is the barrier nuclei need to overcome or tunnel through for fusion to occur. The fusion probability therefore depends on the characteristics of this barrier, such as its height and width. Let us recall that in a star, the energy of reactants is related to their thermal motion: fusion of carbon occurs at center of mass energies in the range $E_{cm} \approx 1-3$~MeV. These energies are lower than the Coulomb barrier for $^{12}$C~+~$^{12}$C fusion reaction, $E_{CB} = 6.6$~MeV.

The first particle accelerators were developed in the 1930s allowing nuclear reactions to be studied for the first time. Such accelerators, however, could only accelerate light ions such as protons and it was not until the 1960s that accelerator technology had developed sufficiently that heavy-ion beams such as $^{12}$C could be accelerated. An early focus with such accelerators was the study of heavy-ion fusion. The corresponding descriptions of the fusion cross sections were based on the transmission of the CB, and on the probability that a compound nucleus --- the composite of the nucleons in the target and projectile --- was formed during the collision. This compound nucleus, formed in a very hot state, would then emit particles (neutron, proton, alpha) subsequently leading to residual nuclei in an excited state that then emit gamma-rays.

In the 1980s, the exploration of fusion below the CB led to the measurement of cross sections which were, surprisingly, orders of magnitude larger than those predicted by simple tunneling through a 1D potential barrier. This was the discovery of heavy-ion fusion enhancement, which recognises that sub-barrier fusion critically depends on intrinsic degrees of freedom of the interacting nuclei: deformation, low-lying collective excitations, as well as the exchange of nucleons \citep{Jacobs1983,Dasgupta1998}. Coupled-channels models, including couplings to all relevant degrees of freedom, subsequently proved extremely successful in describing fusion below the CB. In a coupled-channels calculation, the nuclear potential has an impact on the CB characteristics as well as on the coupling strengths.

In the early 2000s, advances in experimental techniques and detector technology allowed cross sections to be accurately measured at deep sub-barrier energies, even down to cross sections in the nanobarn range \citep{Jiang2002}. A surprising effect found at such low energies is the phenomenon of fusion hindrance, that is an unexpectedly steep fall-off in the fusion cross section systematically below the predictions of coupled-channels calculations using standard potentials. This was first explained either using the saturation property of nuclear matter \citep{Misicu2006} or by adding a second potential barrier which the system of colliding nuclei would have to overcome, after reaching a touching configuration, to finally fuse to form the compound nucleus \citep{Ichikawa2007}. More recently, another explanation was proposed, reminding us that nucleons are fermions and taking into account the Pauli repulsion effects in computing the fusion cross sections \citep{Simenel2018}. The argument that fusion hindrance might affect all nuclear fusion reactions at the lowest energies is still hotly debated. It is thus essential to understand the onset of this phenomenon, which usually occurs at excitation energies for the compound system of $\sim$ 10-20 MeV.
%\\
%– Explanation hindrance vs chemical reactions (David ?) / Nice Figure here ? –
%\\
%{\bf GM: This paragraph is not so clear to me. The link between energy conservation-hindrance - negative Q value is not obvious to me.}

Hindrance has been observed in a large number of heavy-ion fusion reactions, most of them involving medium-mass to heavy nuclear systems. The occurrence of fusion hindrance has been much less explored in lighter systems, especially those of astrophysical interest. Measuring in those systems is therefore of high importance since the relevant energy region for fusion hindrance to occur may correspond to the Gamow window and thus constrain S-factor extrapolations and models \citep{Back2014}.

%\section{Nuclear fusion} \label{sec:nf}
%\subsection{$ ^{12}$C + $^{12}$C fusion reaction}\label{subsec:12C}

\subsection{The STELLA experiment} \label{subsec:se}
%Technical developments (expliqués simplement, Sandrine) \\
%Cross sections well below the nanobarn level. Sensitivity to background.
%\begin{itemize}
%  \item Rotating target system that can sustain high beam intensities
%  \item High efficiency particle and gamma-ray detection system for coincidence measurements, under high vacuum
%  \item Specific analysis to reduce the background (timing) and treatment of low statistics.
%\end{itemize}

%The accurate determination of fusion cross-sections at the nanobarn level requires effective background-suppression methods \citep{Aguilera2006}. Contemporary measurements therefore often employ coincident detection of light charged particles and gamma rays \citep{Jiang2012}. 

% - Comme évoqué dans l'introduction/sect. 1, la mesure de la section efficace de $^{12}$C~+~$^{12}$C rencontre deux principaux obstacles : resonance and background.

The direct measurement of cross sections down to the nanobarn level is very challenging from the point of view of background suppression, the necessity of handling intense beams, and the treatment of data associated with rare events. For example, in such measurements, the gamma-ray signal of interest often becomes significantly less than the environmental gamma-ray background, and minor isotopes such as deuterium in water absorbed into the carbon target foils, produce dominant charged-particle backgrounds which can make discrimination of fusion charged particles highly challenging. Contemporary measurements therefore often employ coincident detection of light charged particles and gamma rays to significantly reduce the type of backgrounds discussed \citep{Jiang2012}.
% - Une solution proposée pour répondre à ce background : la mise en place de la technique de détection en coïncidence proposée par \citet{Jiang2012)}, dont la finalité est de détecter simultanément les particules légères et les rayonnement gammas de déexcitation qui leur sont associés.

An example of a system which employs the coincidence approach is the
STELLA apparatus~\citep{Heine2018}. STELLA exploits several specific technical developments to maximise sensitivity and background rejection, tailored to the challenging application of measuring carbon-fusion at very low cross sections. STELLA combines double-sided silicon strip detectors for charged-particle detection with an array of LaBr$_{3}$(Ce) gamma-ray detectors \citep{Robert2014} that offer excellent timing resolution \citep{Heine2018}. Data are collected using a triggerless data acquisition system. To support the high beam currents required for such studies, a rotating target mechanism is employed to efficiently dissipate heat in the thin self-supporting target foils, allowing beam intensities of several microamps to be safely handled. The stability of the detection system is monitored offline during data-taking periods of months. Accuracy at the lowest counting rates \citep{Feldman1998} is achieved by comparing sparse count measurements with an intrinsic background estimation \citep{Fruet2020}.

\section{Results}

\subsection{Cross sections} \label{subsec:cs}

\begin{table*}[]
\caption{Parameters of cross sections for $^{12}$C~+~$^{12}$C fusion reaction using different models from data interpolation. 
}    
\label{tab:param_cs}
    \centering
    \resizebox{\linewidth}{!}{%
    \begin{tabular}{lcccccccc}
    \hline \hline
    \noalign{\smallskip}
    Model& $\sigma_{s}$& $E_{s}$& $A_{0}$& $B_{0}$& $((\omega \gamma )_{R})_{\alpha}$& $((\omega \gamma ) _{R})_{p}$& $\Gamma _{R}$& $E _{R}$ \\
& (10$^{-2}$ mb)& (MeV)& (MeV$^{-1}$)& (10$^{1}$ MeV$^{1/2}$)& (meV)& (meV)& (keV)& (MeV) \\
    \noalign{\smallskip}
    \hline
    \noalign{\smallskip}
    Hin & 2.20$\pm$0.14 & 3.69$\pm$0.01 & -1.16$\pm$0.03 &  5.13$\pm$0.02 & $\varnothing$ & $\varnothing$ & $\varnothing$ & $\varnothing$ \\
    HinRes & 2.20$\pm$0.14 & 3.69$\pm$0.01 & -1.11$\pm$0.03 &  5.09$\pm$0.02 & 0.11$\pm$0.03 & 0.02$\pm$0.03 & 12 & 2.138$\pm$0.006 \\
    \noalign{\smallskip}
    \hline
    \end{tabular}}
\end{table*}

\begin{figure}[htb!]
\centering
\includegraphics[width = 0.49\textwidth]{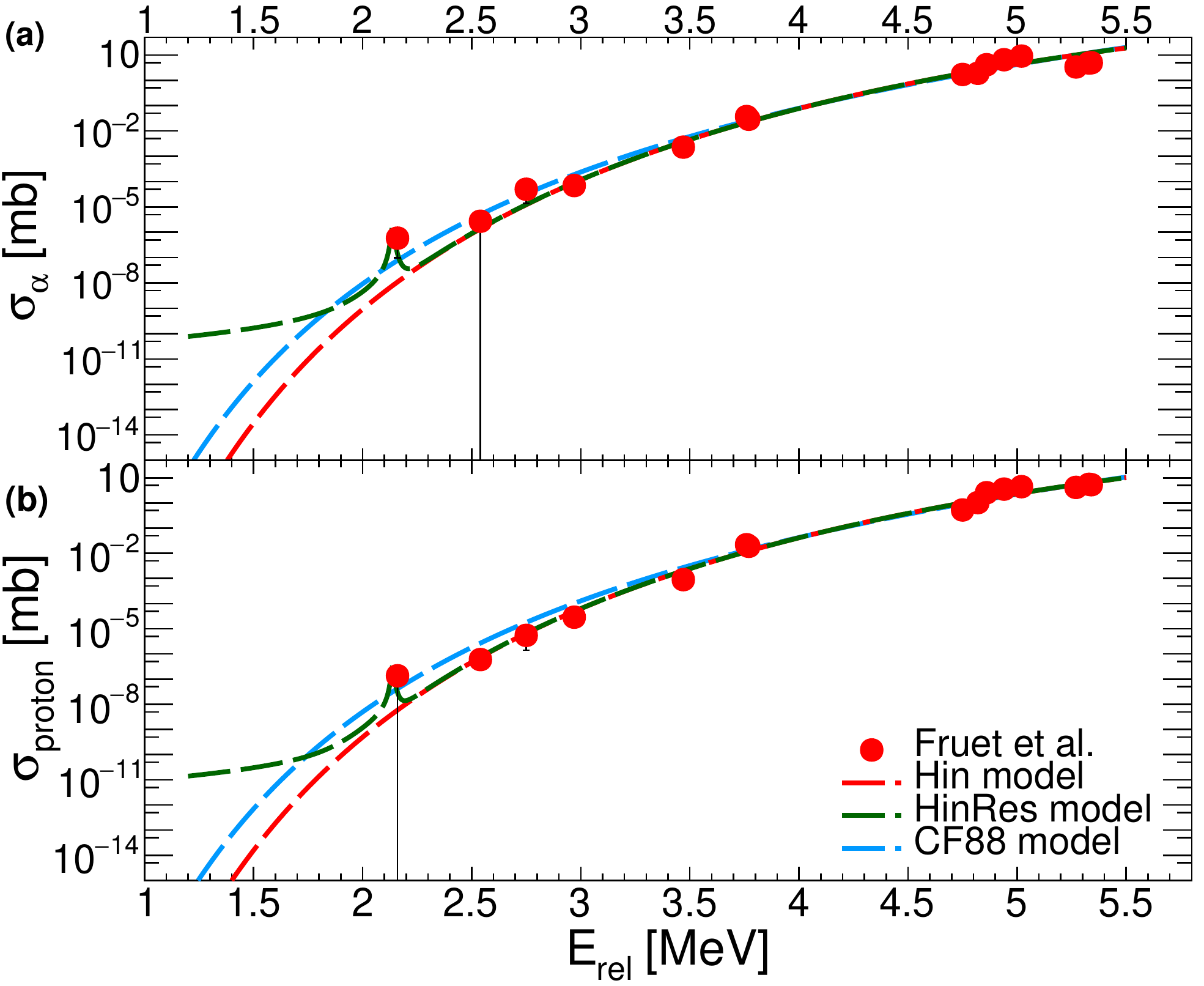}
\caption{Cross sections for alpha (a) and proton (b) channels, obtained with STELLA (red points). Adjustments made for the Hin model (red curve) and HinRes model (green curve) are compared with the cross section from the CF88 model (blue curve). Data points at the lowest energy for protons and the second lowest for the $\alpha$ channel are upper limits (vertical black lines).  \label{fig:cs_c}}
\end{figure}

\begin{figure*}[htb!]
\centering
\includegraphics[width = 0.49\textwidth]{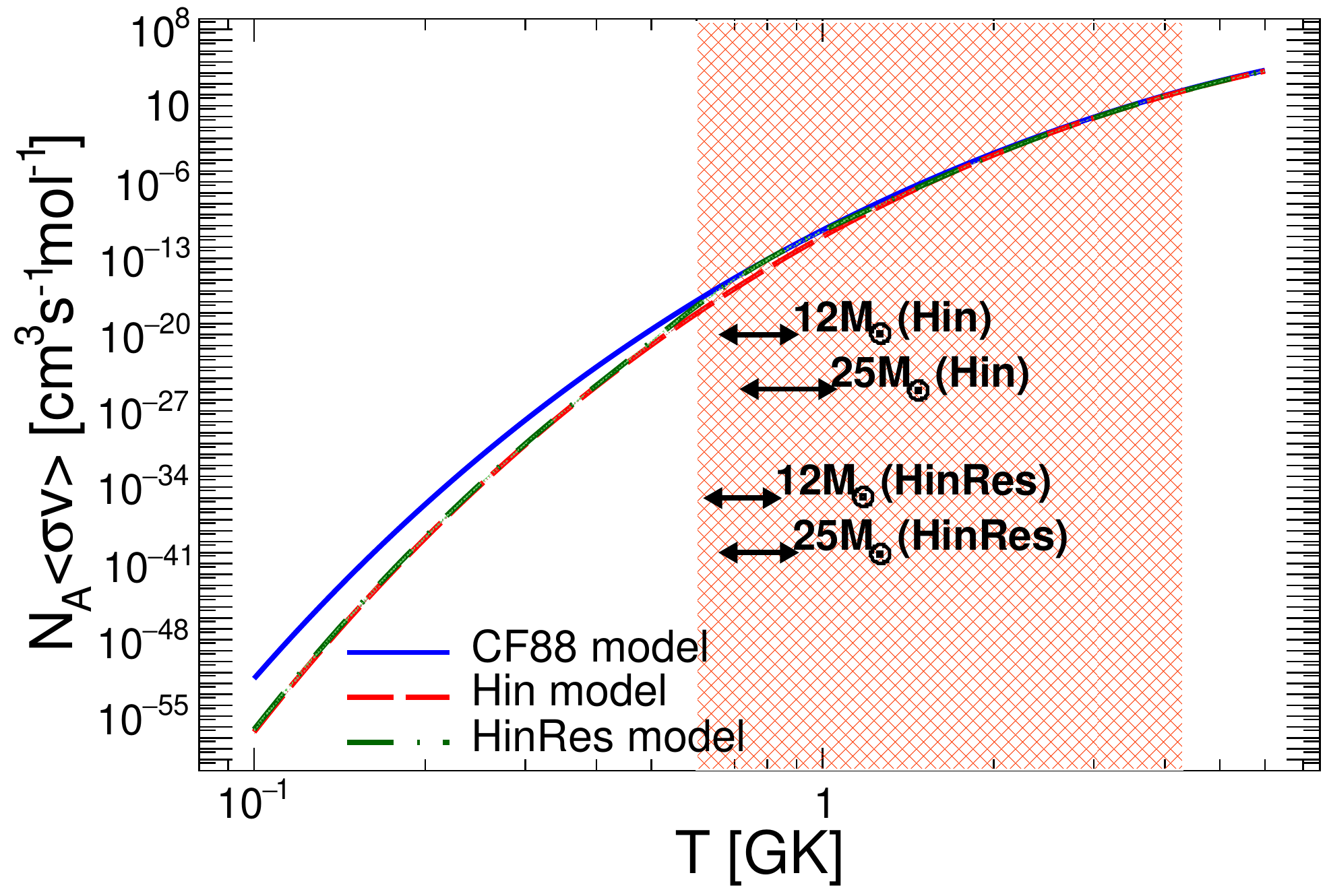}
\includegraphics[width = 0.49\textwidth]{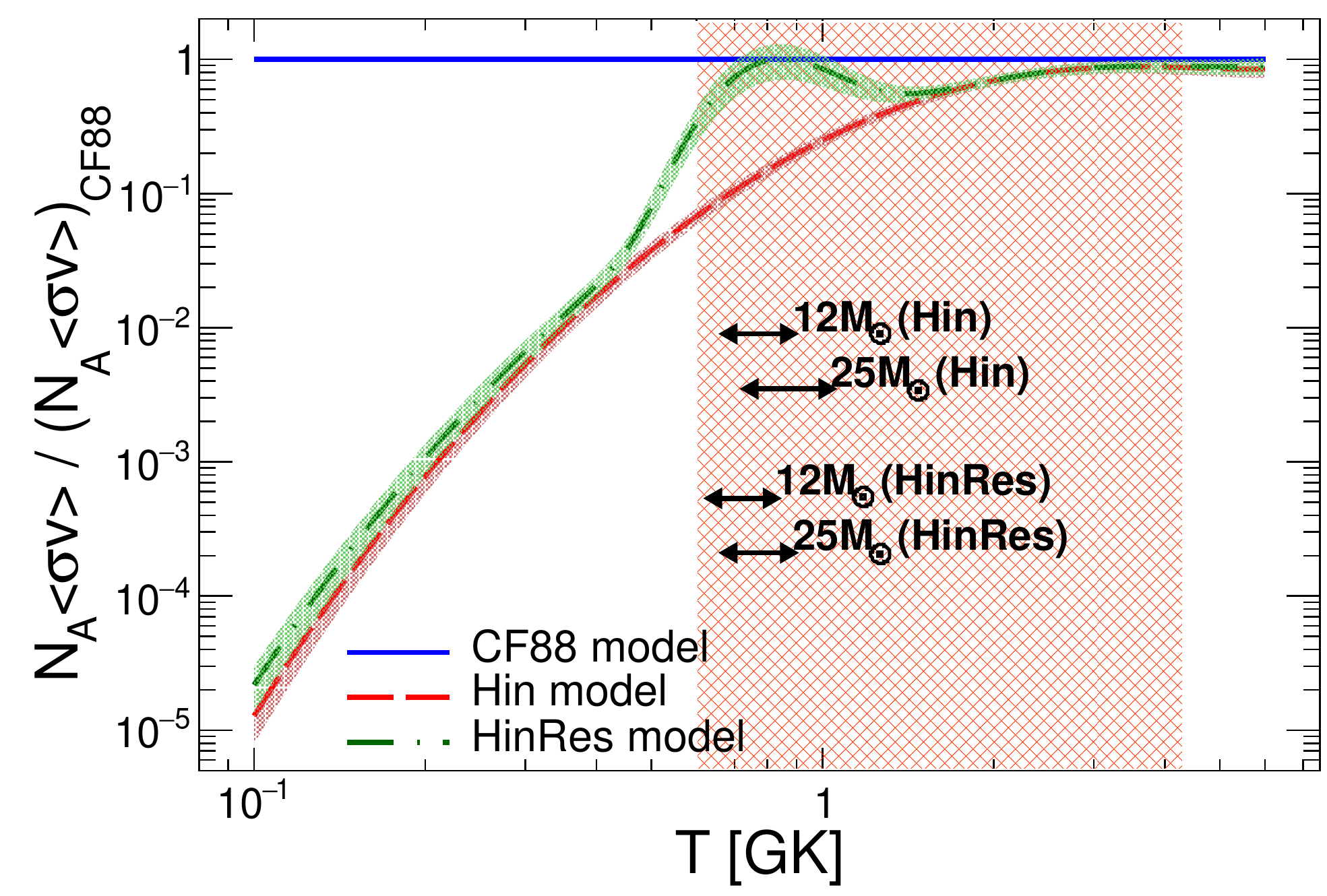}
\caption{Reaction rates (left) and normalized reaction rates on $N_{A} \langle \sigma v \rangle)_{CF88}$ (right), without (Hin; red curve) and with (HinRes; green curve) the added resonance. The reaction rate from the CF88 model is also presented (blue curve). The shaded areas around the curves are the uncertainties (see text). Orange, hatched areas show the temperature region explored by the STELLA experiment. The black arrows show the regions where carbon fusion occurs, for two stellar models (12 and 25 \Msol), for both the Hin and HinRes models (see Sect. \ref{sec:sm} and Fig. \ref{fig:Tc_C12}).
\label{fig:rr}}
\end{figure*}

%Taking account of the different observations carried out for the fusion reaction system, Marcel 2021-07-23

The sub-barrier $^{12}$C + $^{12}$C fusion excitation function exhibits a series of resonances \citep{Aguilera2006}. The present deep sub-barrier data suggest the existence of an intermediate energy region with hindrance trend \citep[see][]{Fruet2020} and an onset of dominant resonance structures towards even lower energies. Therefore, two scenarios have been explored. The first one considers that the cross section follows an empirical model of fusion hindrance (Hin model). The second one is based on the hindrance model, on top of which a resonance proposed by \citet{Spillane2007} has been added (HinRes model). The HinRes model provides a better fit to the measured cross-section data, but the presence of the resonance suggested by the data point at the lowest measured energy may need confirmation from further experiments. The procedure is described in the following paragraphs.

An excitation function for the Hin model has been obtained by fitting data measured by the STELLA collaboration --- \citet{Fruet2020} --- with a dependency suggested by \citet{Jiang2007}:
\begin{eqnarray}
    \sigma(E) _{hin} = \sigma_{s} \frac{E_{s}}{E} \exp  \left ( A_{0} (E-E_{s}) -  \frac{2B_{0}}{\sqrt{E_{s}}} \left (\sqrt{\frac{E_{s}}{E}} - 1 \right ) \right ) \, , 
\end{eqnarray}
where the parameters $E_{s}$ and $\sigma_{s}$ are the center of mass energy and the total cross section, respectively, for which the S-Factor $S(E)$ is maximum, and $A_{0}$ and $B_{0}$ are fitting parameters. All parameters are free during the adjustment of $\sigma(E) _{hin}$ to STELLA data. The formula describing the total cross section was scaled by branching ratios from previous experimental data \citep{Pignatari2013}, corresponding to 35\% for the proton exit channel and 65\% for the alpha exit channel.

In the second scenario (HinRes), a resonance is added on top of the Hin model at the lowest energy; the contribution from an isolated narrow resonance described by the one-level Breit-Wigner cross section formula is:
\begin{equation}
    \sigma_{BW}(E) =  \frac{\lambda ^{2}}{4 \pi}  \frac{(2 J + 1)(1 + \delta _{01})}{(2 j_{0} + 1)(2 j_{1} + 1)} \frac{\Gamma _{a} \Gamma _{b}}{ (E_{R} - E )^{2} + \Gamma _{R} ^{2} / 4}  ,
\end{equation}
where $j_{0}$ and $j_{1}$ are the spins of target and projectile respectively, $J$ and $E_{R}$ are the spin and the center of mass energy of the resonance, $\Gamma _{a}$ and $\Gamma _{b}$ are the resonance partial widths of entrance and exit channel, $\Gamma _{R}$ is the total resonance width, and $\delta _{01}$ is the Kronecker delta. Values of the resonance parameters are kept fixed throughout the rest of this work. The cross section considered for this second scenario is the sum of Hin model cross section and the Breit-Wigner cross section. 

The parameters were adjusted in both exit channels simultaneously in the range covering the expected hindrance maximum in the fitting procedure. Values of these parameters are given in Table \ref{tab:param_cs}. They are consistent with those predicted by \citet{Jiang2007}, with a cross section maximum at energy $E_{s} = 3.69$ MeV.
%For both scenarios, we performed adjustments simultaneously on all exit channels, for different energies in the range $E_{eff} = 2.54 - 4.86$ MeV. The effective energy $E_{\rm rel}$ is the center of mass energy to which the reaction occurs at the target center \citep{Fruet2020}. 

Cross sections and interpolations are shown in Fig.~\ref{fig:cs_c}: in red for the Hin model and in green for the HinRes model. Cross sections predicted by \citet{CF88} (CF88 model) are also presented in blue. Cross sections have been extrapolated here down to an energy of $E_{rel}=1.2$ MeV.

We observe that at energies near the Coulomb barrier (i.e. $E_{CB} = 6.6$ MeV), the models overlap, but diverge towards the lower energies, where the Hin model is slightly below that of the CF88 model. The resonance predicted by \citet{Spillane2007} shows a good compatibility with the STELLA measurements at the resonance energy. For energies below $E_{rel} = 2$ MeV, we note that while the cross section extrapolations from the Hin model and the CF88 model follow the same trend, the cross section from the HinRes model follows the tail of the resonance.

\subsection{Reaction rates} \label{subsec:rr}

In a general way, the stellar reaction rate $(N_{A} \langle \sigma v \rangle)$ can be expressed as \citep{Rolfs}:
\begin{equation}
\label{eq::rr_hin}
    (N_{A} \langle \sigma v \rangle) = \left ( \frac{8}{ \pi \mu} \right )^{1/2} \frac{N_{A}}{(kT)^{3/2}} \int _{0}^{\infty} \sigma(E) E \exp \left (- \frac{E}{kT} \right ) dE,
\end{equation}
where $N_{A}$ is Avogadro's number, $\mu$ is the reduced mass of the system, $k$ is the Boltzmann constant, $T$ is the temperature at which the reaction occurs, $\sigma(E)$ is the reaction cross section, and $E$ is the center of mass energy. The total cross section $\sigma (E)$ is obtained by summing the cross section from all exit channels.

In the HinRes scenario, the reaction rate is determined by adding a rate specific to a single narrow resonance to the one described in Eq. \ref{eq::rr_hin}. \citet{Illiadis} define it as:
\begin{eqnarray}
    (N_{A} \langle \sigma v \rangle)_{res_{i}} = \frac{1.5399 \times 10^{11}}{ \left ( \frac{M_{0}M_{1}}{M_{0} + M_{1}} T_{9} \right )^{3/2} } \sum_{i} \left ( (\omega \gamma)_{R} \right ) _{i} \exp \left (-11.605 \frac{(E_{R})_{i}}{T_{9}} \right ) , \nonumber
\end{eqnarray}
\vspace*{-0.7cm}
\begin{eqnarray}
    &
\end{eqnarray}
where \textit{i} labels each resonance, $\left ( (\omega \gamma)_{R} \right ) _{i} $ and $E_{R}$ are the resonance strength and energy, $M_{i}$ is the reactant atomic mass, and $T_{9}$ is the temperature (in GK) at which the reaction occurs.

%The reaction rates obtained are shown in Fig. \ref{fig:rr}, in red for the Hin model and green for the HinRes model. The stellar reaction rate predicted in \citet{CF88} $(N_{A} \langle \sigma v \rangle)_{CF88}$ (CF88 rate) is also presented in blue. The orange hatched areas show the temperature region explored by the STELLA measurements. The shaded areas around each curve show the total uncertainties in the reaction rates, determined through error propagation of experimental uncertainties in the STELLA cross sections. The temperature regions where C-burning occurs are also shown on Fig. \ref{fig:Tc_C12}, for both stellar models studied in Sect. \ref{sec:sm}, and for both scenarios considered in this work. 
The reaction rates obtained are shown in Fig. \ref{fig:rr}, in red for the Hin model and green for the HinRes model. The stellar reaction rate predicted in \citet{CF88} $(N_{A} \langle \sigma v \rangle)_{CF88}$ (CF88 rate) is also presented in blue. The orange hatched areas show the STELLA sensitivity, that is the temperature region where no extrapolation of the cross sections measured by STELLA is required. The shaded areas around each curve show the total uncertainties in the reaction rates, determined through error propagation of experimental uncertainties in the STELLA cross sections. %The temperature regions where C-burning occurs, {\bf determined with data from Fig. \ref{fig:Tc_C12},} are also shown, for both stellar models studied in Sect. \ref{sec:sm}, and for both {\bf nuclear fusion} scenarios considered in this work. 
%The {\bf central temperature during C-burning phase ({\it i.e.} the last model before the central carbon abundance is lower than 10$^{-5}$), determined with data from Fig. \ref{fig:Tc_C12},} are also shown, for both stellar models studied in Sect. \ref{sec:sm}, and for both {\bf nuclear fusion} scenarios considered in this work.
The central temperature during C-burning phase (which begins here when 1\% of the carbon abundance in the core is consumed, and ends when this same abundance is less than 10$^{-5}$), determined with data from Fig. \ref{fig:Tc_C12}, are also shown, for both stellar models studied in Sect. \ref{sec:sm}, and for both nuclear fusion scenarios considered in this work.

In the Hin scenario, the results obtained show that the reaction rate from STELLA data and the hindrance model is lower than that predicted by the CF88 model at low temperatures, but is similar to it at high temperatures, which is consistent with the predictions of \citet{Jiang2007}.

In the HinRes scenario, the reaction rate is essentially the same as the one without a resonance, and it persists lower that the one predicted in CF88. However, we can see that the resonance has two main effects on it. It slightly increases the reaction rate at low energies, and, in an intermediate region between $T=0.5$ and 1.5~GK, it significantly increases the reaction rate to a level comparable with the CF88 rate.

%The lowest center of mass energy reached by the STELLA experiment is $E=2.03$ MeV \citep{Fruet2020}, corresponding to the Gamow energy for a temperature, $T=0.77$ GK. We define this value as the sensitivity limit of the present data set. In the temperature range, $T = 0.77 - 3.33$ GK, the relative uncertainty for the Hin model is below 13~\%, and reaches 30~\% in the vicinity of temperatures where the resonance has the most important impact for the HinRes model. This difference may be explained by the large uncertainties on parameters describing the resonance.  

The lowest centre-of-mass energy reached in the STELLA experiment is $E = 2.03$~MeV (see supplements in Fruet et al. 2020), corresponding to the Gamow energy at $T = 0.77$~GK. In the present case, the Gamow window can be very well approximated by a Gaussian function with the left side sigma-interval, that is a $68.27\%$ probability coverage, translating into a lower temperature $T=0.60$~GK. We associate this value to the sensitivity limit of the present experimental fusion-measurement data for detecting either a Hin or a HinRes trend, within the commonly used one-sigma significance-level.

In the temperature range $T = 0.6 - 4.3$~GK, the relative uncertainty of the fits for the Hin model is below 15$\%$ and only becomes 31$\%$ towards lower temperatures in the case of the HinRes trend, due to the large uncertainties of the resonance parameters.

\section{Stellar Models} \label{sec:sm}
\defcitealias{CF88}{CF88}

\begin{figure*}
    \centering
    \includegraphics[width=0.85\textwidth]{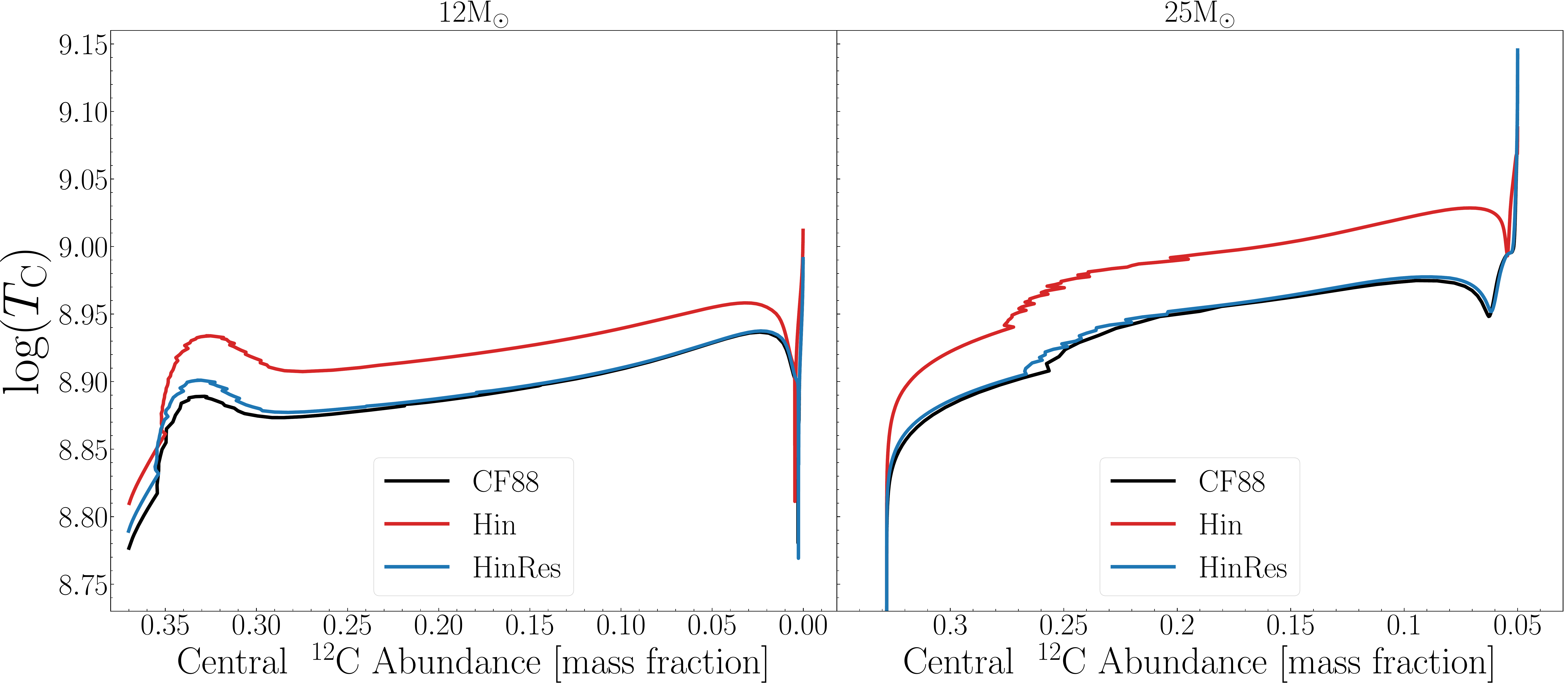}

    \caption{Central temperature evolution during the C-burning phase for 12\Msol\ and 25\Msol\ models with different $^{12}$C~+~$^{12}$C reaction rates. The evolution is given as a function the mass fraction of carbon at the centre that decreases as a function of time.}
    \label{fig:Tc_C12}
\end{figure*}

% \begin{figure*}
%     \centering
%     \includegraphics[width=0.3\textwidth]{figures/Kippen_nuc0.pdf}
%     \includegraphics[width=0.3\textwidth]{figures/Kippen_nuc1.pdf}
%     \includegraphics[width=0.305\textwidth]{figures/Kippen_nuc2.pdf}
%     \caption{Kippenhan diagrams for 25\Msol\ models during the central C-burning phase with different $^{12}$C+$^{12}$C reaction rates. The grey shaded area shows the convective zones. The blue, green and red dashed line show respectively the limits of the H-burning, He-burning and C-burning zones.}
%     \label{fig:Kippenhan Diagrams}
% \end{figure*}

\begin{figure*}
    \centering
    \includegraphics[width=0.44\textwidth]{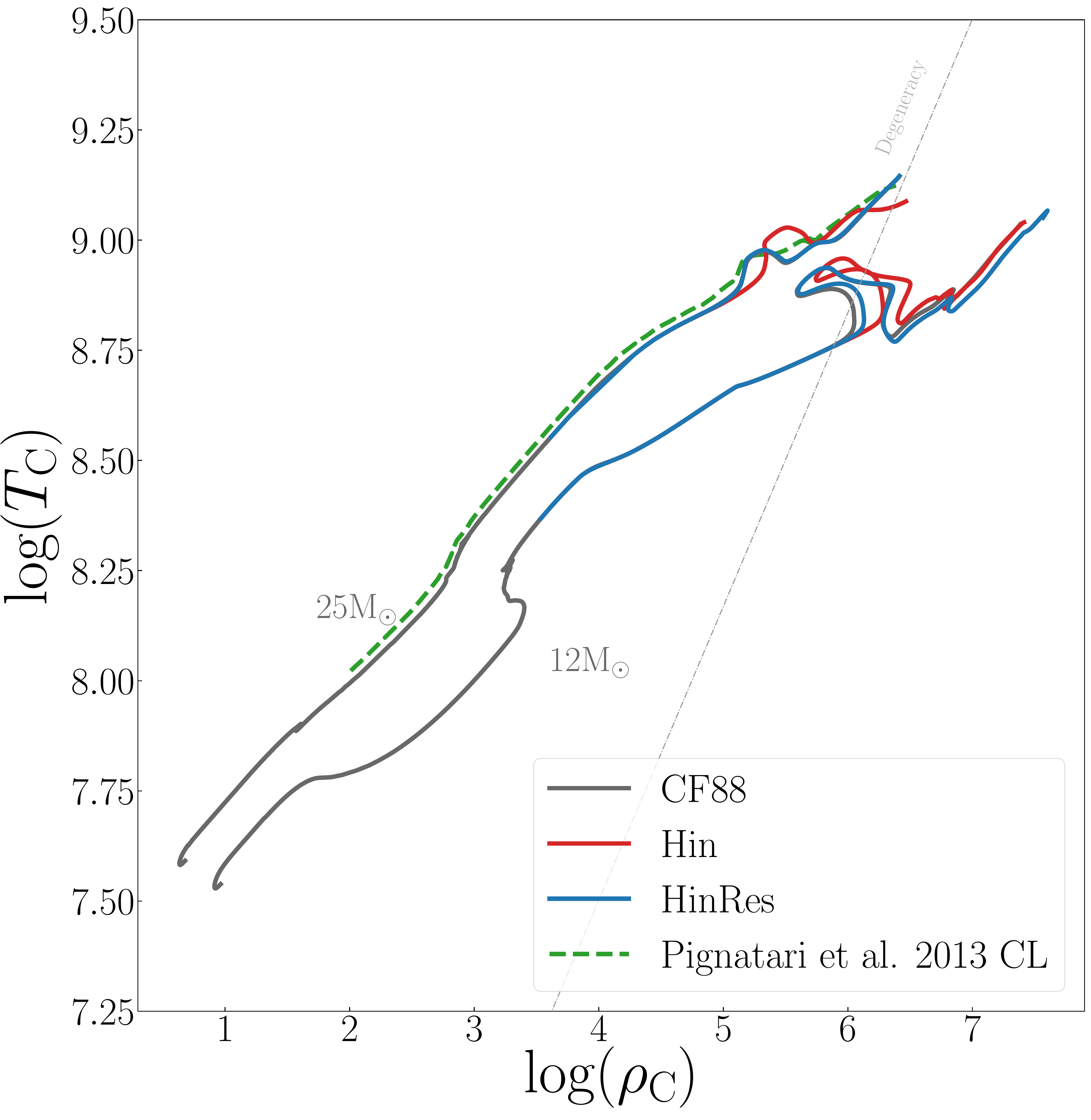}\hfill\includegraphics[width=0.45\textwidth]{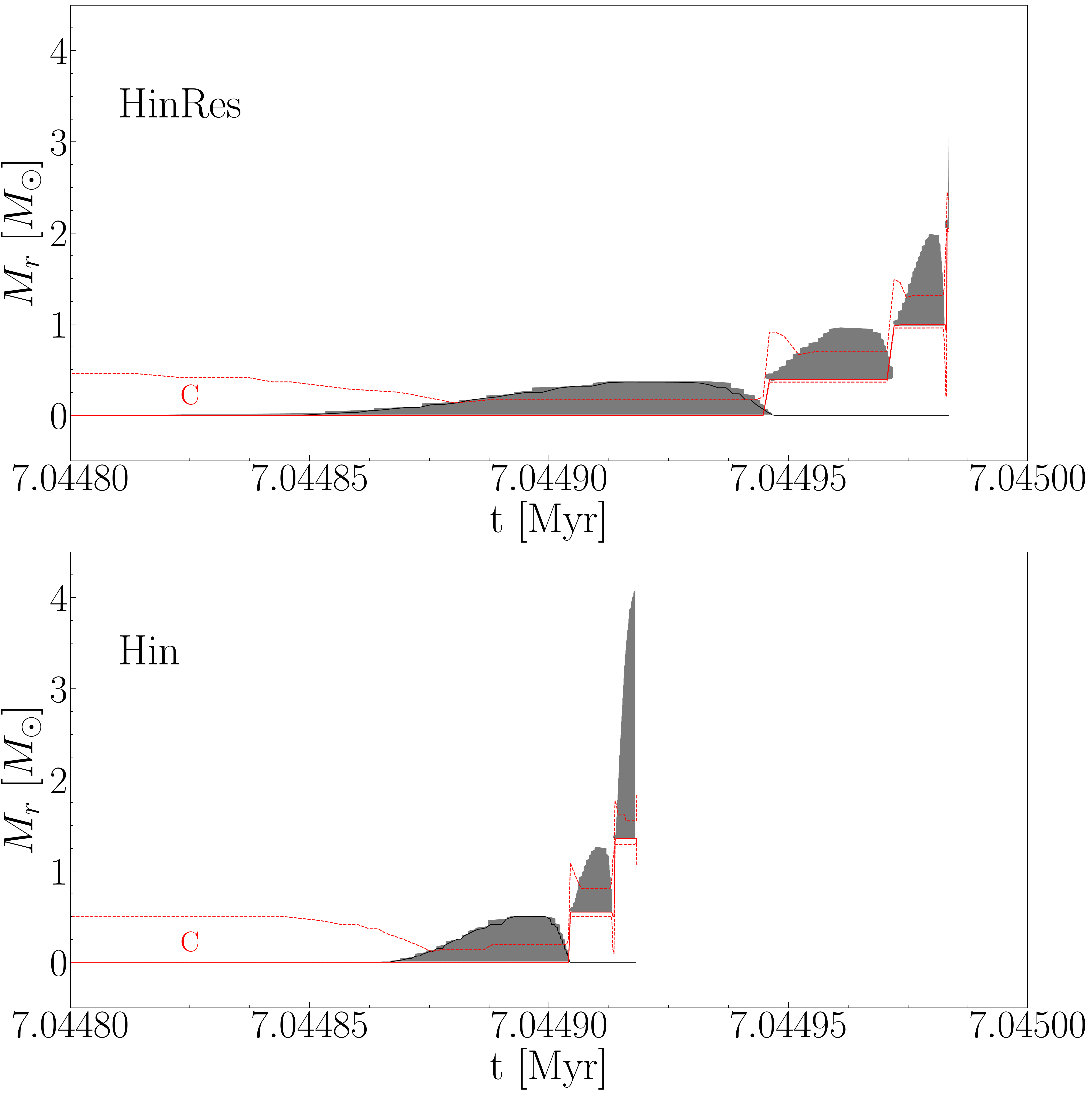}
    \caption{{\it Left panel:} Evolution of central temperature as a function of central density for 12\Msol\ and 25\Msol\ models with different $^{12}$C~+~$^{12}$C reaction rates. {\it Right panel:} Kippenhahn diagrams for the centre of 25\Msol\ models during the end of C-burning phase with Hin and HinRes $^{12}$C~+~$^{12}$C reaction rates. The grey shaded area shows the convective zones. The red dashed line show  the limits of the C-burning zones (defined where $\epsilon_{\rm C}\geq 10^2$erg g$^{-1}$s$^{-1}$).}
    \label{fig:rhoT}
\end{figure*}

\begin{figure*}
    \centering
     \includegraphics[width=0.45\textwidth]{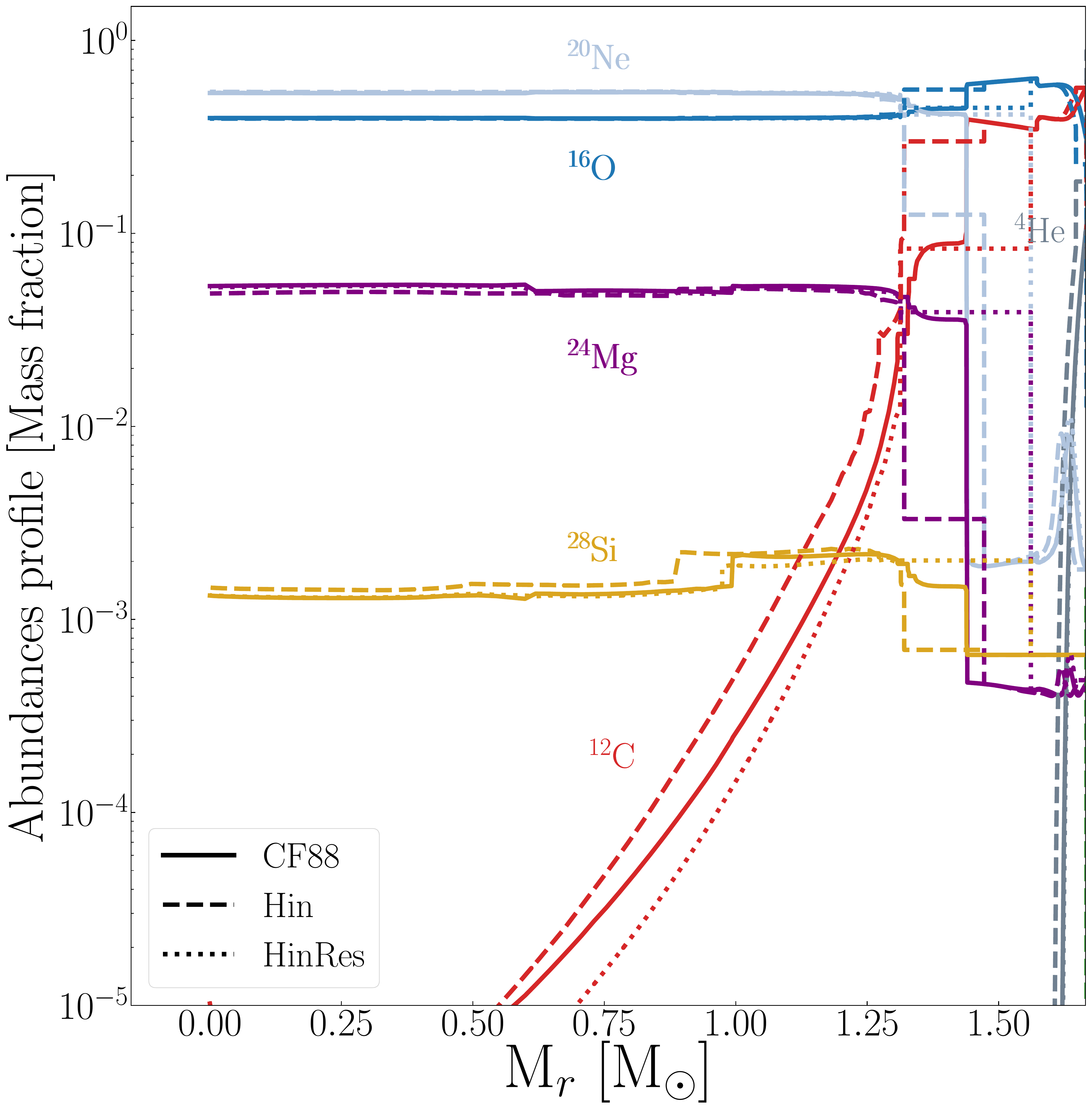}       \includegraphics[width=0.45\textwidth]{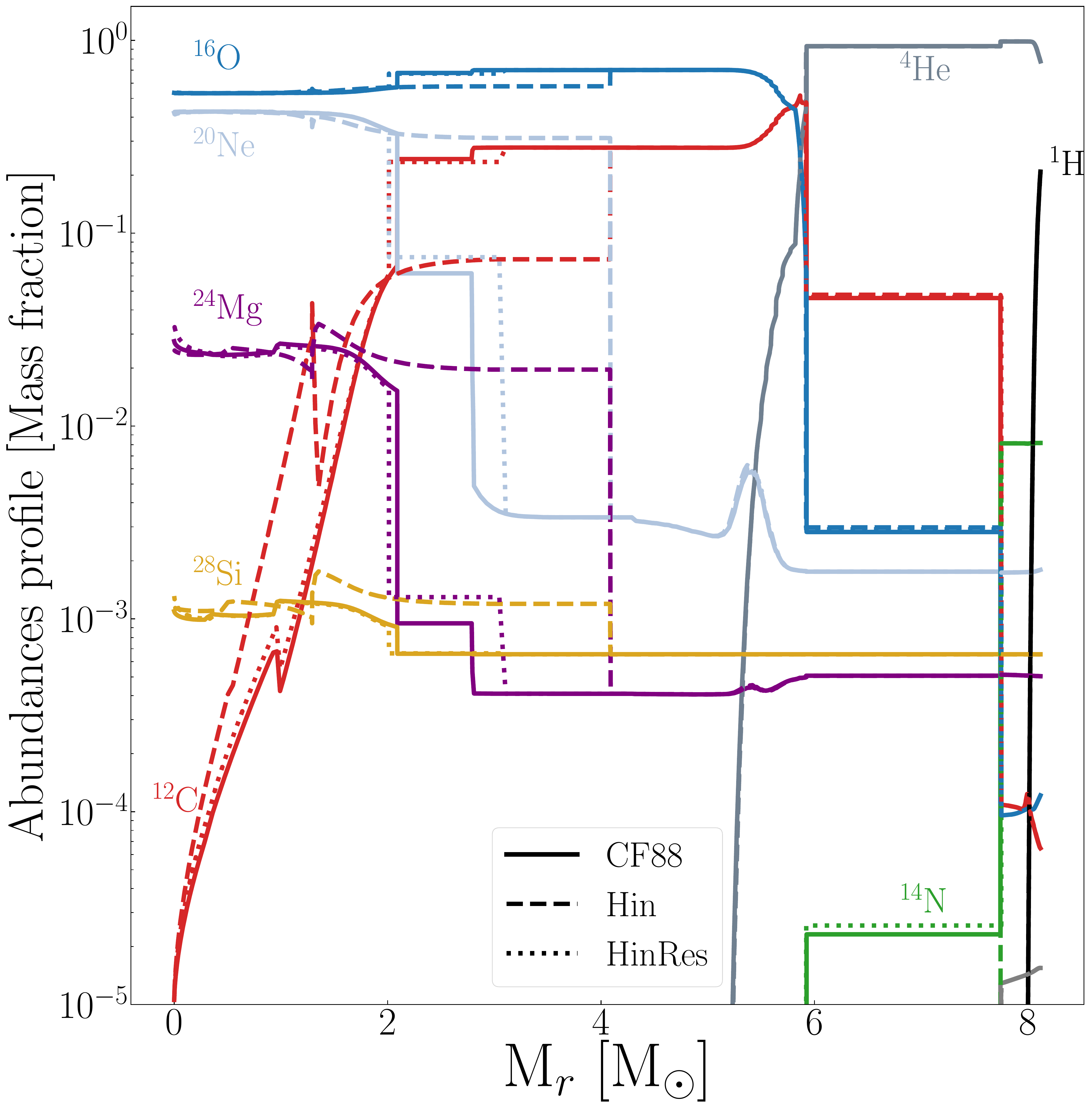}
    \caption{{\it Left panel:} Core abundances profile at the end of central C-burning for 12\Msol\ models with different $^{12}$C~+~$^{12}$C reaction rates. {\it Right panel:} Abundances profile at the end of C-burning for 25\Msol\ models with different $^{12}$C~+~$^{12}$C reaction rates.}
    \label{fig:abudances_12}
\end{figure*}

% \begin{figure}
%     \centering
%     \includegraphics[width=0.8\textwidth]{figures/Abundances_profile_25Msol_whole_star.pdf}
%     \caption{Abundances profile at the end of C-burning for 25\Msol\ models with different $^{12}$C+$^{12}$C reaction rates.}
%     \label{fig:abudances_25}
% \end{figure}

%\begin{figure*}
%    \centering
%    \includegraphics[width=0.8\textwidth]{figures/Abundances_triple_plot_total.pdf}
%    \caption{\textit{Left: }Abundances profile at the end of central C-burning for 25\Msol\ models with different $^{12}$C+$^{12}$C reaction rates. \textit{Right:} Total and relative integrated abundances over the whole structure for the same models that on the left panel.}
%    \label{fig:abudances_25}
%\end{figure*}

%\begin{figure}
%    \centering
%    \includegraphics[width=0.45\textwidth]{Kippenhan_zoom.pdf}
%    \caption{Kippenhan diagrams for the center of 25\Msol\ models during the end of C-burning phase with hindrance and hindrance plus resonance $^{12}$C+$^{12}$C reaction rates. The grey shaded area shows the convective zones. The red dashed line show  the limits of the C-burning zones.}
%    \label{fig:Kippenhan Diagrams}
%\end{figure}

\begin{figure*}
    \centering
       \includegraphics[width=1.00\textwidth]{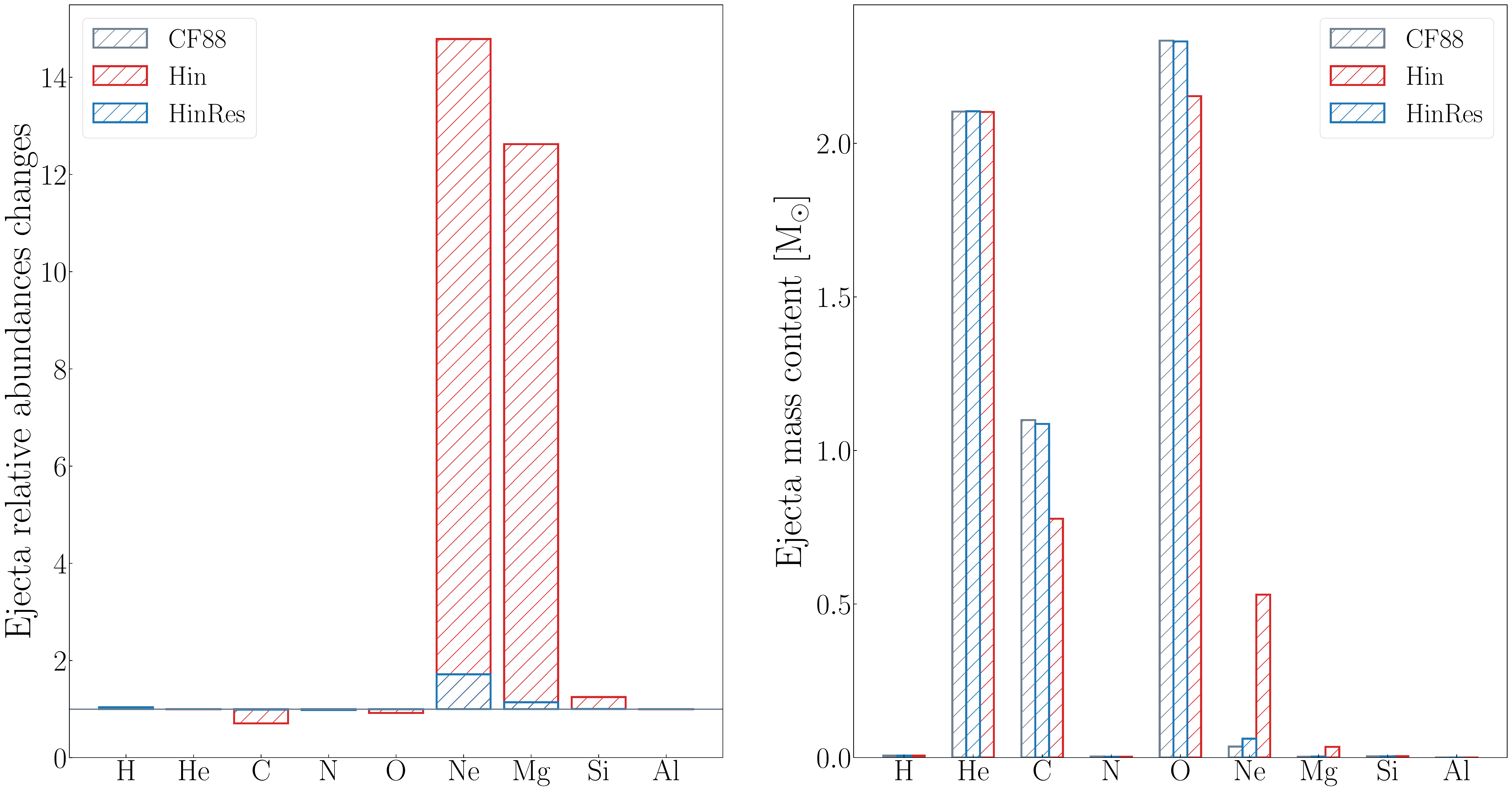}
 \caption{\textit{Left:} Abundances relative to CF88 models in the supernova ejecta of the 25\Msol\ models for different $^{12}$C~+~$^{12}$C reaction rates (based on the structure obtained at the end of the C-burning phase). \textit{Right:} Same but for the total mass content in the ejecta.}
    \label{fig:ejecta_content}
\end{figure*}

\begin{table}[]
    \centering
    \caption{Central C-burning lifetime for different $^{12}$C~+~$^{12}$C reaction rates.}
\label{table:lifetimes}
\begin{tabular}{lccc}
    \hline \hline
    \noalign{\smallskip}
     $^{12}$C~+~$^{12}$C & CF88 & Hin & HinRes\\
    \noalign{\smallskip} 
    \hline
    \noalign{\smallskip}
    12\Msol & 7604 yrs & 3856 yrs& 6698 yrs \\
    25\Msol & 820 yrs& 403 yrs& 717 yrs\\
    \noalign{\smallskip}
\hline
\end{tabular}
\end{table}

\begin{figure}
    \centering
    \includegraphics[width=0.4\textwidth]{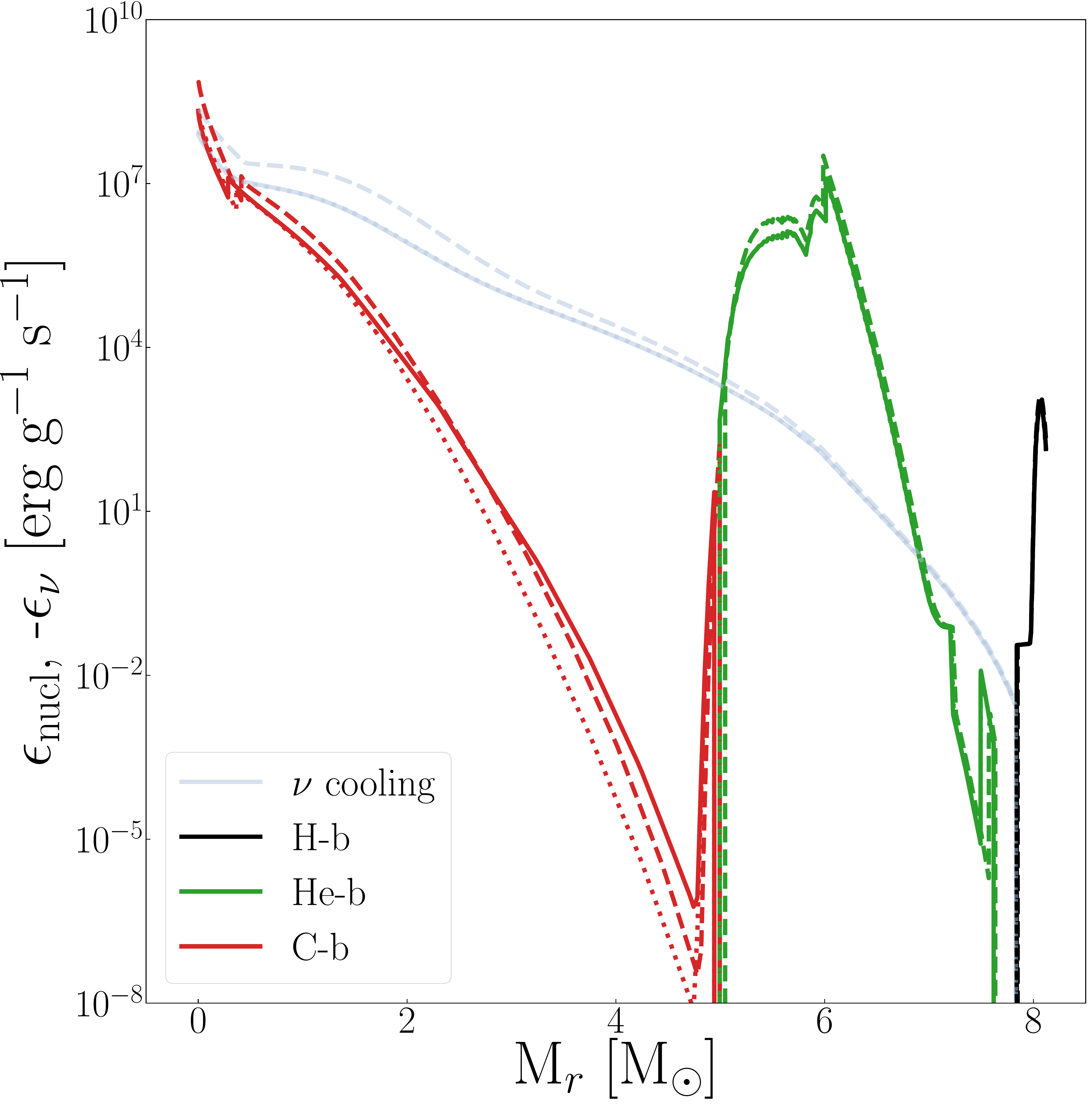}
    \caption{Energy production at the mid-point of C-burning (when half the central $^{12}$C has been consumed since the beginning of C-b) for 25\Msol\ models for different $^{12}$C~+~$^{12}$C reaction rates. The CF88, Hin and HinRes models are respectively plotted in plain, dashed and dotted lines.}
    \label{fig:energy_production_25}
\end{figure}

In this section, we explore the impact of the results obtained in Sect.\ref{subsec:rr} on stellar evolution models. Using GENEC \citep{Eggenberger2008}, we computed three sets of models using, respectively, $^{12}$C~+~$^{12}$C reaction rates from \citet[]{CF88} (which we refer to below as CF88), the Hin model, and the HinRes model developed in the present work. For each of these sets, we computed the evolution of a non-rotating 12\Msol\ and 25\Msol\ star at Z=0.014 (solar-like) until the end of the core C-burning phase. For a given initial mass, all the models share the same evolution up until the end of He-burning.
Other than the $^{12}$C~+~$^{12}$C reaction rates, the physics inputs used for these models are the same as those described in \citet[]{Ekstrom2012}.

Figure \ref{fig:Tc_C12} shows the evolution of central temperature during the C-burning phase for different $^{12}$C~+~$^{12}$C reaction rates. 
%Central temperatures are between 0.7 and 1.0 GK in the Hin model, thus in a range where they differ the most from the rates given by \citet{CF88}. 
%The hindrance model shows the largest differences in central temperatures with respect to the \citetalias{CF88} rates. 
 The models with a lower rate (Hin model) burn carbon at a slightly higher central temperature (the relative increase is about 10\%). The fact that lowering a rate produces an increase of temperature is due to the following well known fact: a lower rate implies less energy released per unit time. On the other hand, the energy lost per unit time at the surface of the star is not mainly determined by the nuclear reaction but by the hydrostatic and radiative equilibrium. To compensate for these losses at the surface when the nuclear energy production is less efficient, the core contracts in order to achieve a new hydrostatic/radiative equilibrium. 
 The contraction increases the central temperature and thus the nuclear reaction rates and hence the energy released per unit time. Note that, since the dependence of the nuclear reactions on temperature is very high for the carbon fusion reaction (typically the rates depend on $T^{27}$), a small
 increase suffices to restore the equilibrium.
 
 %This behavior illustrates a well known feature of how self-gravitating systems at hydrostatic equilibrium behave, in a way, as negative specific heat systems. Here the rate is lower, less energy is produced per unit time, the central region then contracts until the rate of energy production reaches a new radiative equilibrium. This induces the central regions to get warmer. As a result, an initial deficit in the energy production produces, in fact, an increase in temperature. 

The situation with the HinRes model is not much different from the case with \citetalias{CF88} rate.
%In comparison to the \citetalias{CF88} reaction rates, the relatively lower reaction rates due to the hindrance phenomenon mean that the star reaches a higher central temperature at the same stage of evolution. However, the hindrance plus resonance model shows a much more similar behavior to the \citetalias{CF88} rates. 
The resonance decreases the maximum temperature where the hindrance effect would be non-negligible, and as seen in Fig. \ref{fig:rr}, the rate at the peak of the resonance is of the order of the \citetalias{CF88} rates. We see here that in stellar conditions during C-burning, the central temperature is in a range where the presence of the resonance is effective, dampening the hindrance impact. Hence, the hotter medium for the 25\Msol\ model shows even less differences than the 12\Msol\ one between the \citetalias{CF88} rates and the HinRes rates.

The evolution of central conditions is shown in the temperature/density diagram in the left panel of Fig. \ref{fig:rhoT}. The core C-burning phase occurs 
when the bump occurs along the tracks that is around 0.8 GK (Log T=8.9).
The bump in the 12\Msol\ track is particularly well developed. This results from the fact that carbon-ignition occurs in a medium mildly degenerated (the
grey straight lines give the positions in the plane where the perfect gas pressure is equal to the non-relativistic electronic degenerate pressure).
As seen in Fig. \ref{fig:rhoT}, the Hin model show bumps that are shifted to slightly higher densities/temperatures with respect to the other models. This results from the fact that these models have more compact cores during the C-burning phase.
%will tend to go to a higher temperature to compensate for the lower reaction rates, while the HinRes model evolves in a very similar fashion to the one using the \citetalias{CF88} rates. 
For comparison, the 25\Msol\ model using the ``lower limit'' from \citet[][]{Pignatari2013} is superimposed in green. This model uses rates close to our HinRes one and consistently
follow a very similar path as our HinRes model. 
%This model again shows no significant difference from the \citetalias{CF88} model and our HinRes model.
%In summary, the hindrance model does not show any major changes to in the evolution of conditions at the stellar core.

The right panel of Fig. \ref{fig:rhoT} shows the Kippenhahn diagrams during the end of C-burning for the 25\Msol\ models. It shows that the C-burning region (red, dashed lines) evolves in the same way for the different reaction rates. However, the convective zones developing in the inner regions of the star at the end of C-burning, when Ne-burning sets up, are quite different for the Hin model. Indeed, the convective zone, fueled by shell C-burning, extends much further away in the Hin model (see lower panel) compared to the HinRes model (upper panel). This effect is not present in the 12\Msol\ models, hence their Kippenhahn diagrams are not shown here. The largest convective zones in the Hin 25\Msol\ model are likely due to the fact that, as already mentioned above, C-burning occurs at higher temperatures due to the lower nuclear energy generation rate in this temperature range. A stronger temperature gradient builds up that favors the occurrence of a larger convective zone. 
%when the rates are decreased. This favors convection. 
The situation in the 12\Msol\ model is different due to the fact that we are in a medium that is more affected by degeneracy.

%due to the presence of a larger convective envelope inhibiting the extension 
%The extension of the convective shell of C-burning in the Hin model leads to very different results if we now look at the abundances profiles in the models. 

Figure \ref{fig:abudances_12} compares the abundance profiles of the 12\Msol\ and 25\Msol\ models for different $^{12}$C~+~$^{12}$C reaction rates obtained at the end of the core C-burning phase (i.e. the last model before the central carbon abundance is lower than 10$^{-5}$). 
On the left panel, only the central 1.75\Msol\ are shown for the 12\Msol\ model, above this regions, the models present no differences between them. In the right panel, the whole structure of the 25\Msol\ is shown.
%There is no significant enrichment at the end of C-burning for the 12\Msol\ models. 
%Looking at the 12\Msol\ model, there is no significant enrichment at the end of C-burning. 
%The envelope was not shown in Fig.\ref{fig:abudances_12} due to the different models being extremely similar. 
%Now looking at Fig.\ref{fig:abudances_25}, we can see that there is a slight enrichment in carbon for the hindrance model in the core. 
We see that the chemical structure at the end of the C-burning is not much affected by the changes of the rates in the 12\Msol\ model. We note, however, some differences between 1.3 and 1.7\Msol. The mass coordinate at which the carbon abundance changes abruptly (around 1.44\Msol) shifts outwards by about 0.1\Msol\ passing from the Hin model to the CF88 model and again passing from CF88 to the HinRes model. This abrupt change in abundance is due to presence of a convective C-burning shell.
These differences, although non-negligible, are not very significant when considering
other uncertainties linked to the treatment of convection in stellar models \citep[see e.g.][]{Gabriel2014,Kaiser2020}.

%The region in the 25\Msol\ model, corresponding to the one just discussed above in the 12\Msol\ one, is between the mass coordinates 2 and 4\Msol.
We observe also a difference in the positions of the sharp change of the carbon abundance in the 25\Msol\ models depending on the rates used. For the Hin model, the step occurs at about 4\Msol, while for the CF88 and HinRes model, the step occurs around 2\Msol. We have here that, in Hin model, the step is shifted outwards with respect to the other two models, while, as seen above, it was shifted inwards in the 12\Msol. This difference between the 12 and the 25\Msol\ models comes from the extension of the convective C-burning shell, that is slightly more extended for the CF88/HinRes models in the 12\Msol\ than the Hin one, while it is largely extended in the 25\Msol\ Hin model compared to the CF88/HinRes ones. Indeed, the size of the convective regions strongly impact the chemical structure by imposing flat chemical gradients in them. The shift outwards of the sharp carbon abundance step in the 25\Msol\ Hin model is linked to the extension of the last intermediate convective zone that we can see in the bottom panel of Fig.~\ref{fig:rhoT}. This convective shell indeed extends up to around 4\Msol, while its maximum outwards extension in the HinRes model is limited to around 2\Msol\ (see the upper panel of Fig.~\ref{fig:rhoT}).

These two different behaviours for the 12 and 25\Msol\ models illustrate the complex non-linear behaviour of the stars whose evolution results from many tightly intertwined processes. 
%In the 25\Msol\ model, a change in a nuclear reaction rate has an impact on the sizes of the convective regions. 

%The abundances in the 25\Msol\ at the end of the central carbon burning are very similar for all three models except in the region between $\sim$3 and $\sim$4\Msol. This is a consequence of the sizes of the last intermediate convective zone associated to the carbon-burning shell shown in the right panel of Fig.~\ref{fig:rhoT} that is significantly different between the Hin and HinRes models (larger in the former model). 
%If we compare the total amounts of different elements present in the star at that stage, the most affected elements are C (amount decreased by about 25\% in the hindrance model), Ne, Mg and Si that show larger amounts by respectively 60, 75 and 25\%. 
%As we have seen in Fig \ref{fig:Kippenhan Diagrams}, the convective shell fueled by C-burning  extends much further out. 
%This leads to the hindrance model showing enrichment in $^{20}$Ne, $^{24}$Mg and $^{28}$Si. 
%This is quantified in the right-hand panels, showing both the total and relative integrated abundances over the whole structure. For the hindrance model, we see a relative enrichment from the \citetalias{CF88} rates of 75\% for $^{20}$Ne and $^{24}$Mg, and 25\% for $^{28}$Si. 
%The HinRes model shows however no difference compared to the \citetalias{CF88} model.

One can wonder whether these differences in the chemical structure
have an impact of the total quantity of an element in the outer envelope of the star. At the end of the C-burning phase the envelope has nearly reached its final chemical composition because the
subsequent nuclear phases will occur in regions well below this envelope and the time that remains before the explosion is so short that
any changes in the envelope have no time to be significant. Moreover, the explosive nucleosynthesis will affects mainly layers near the core and the abundances of elements around the iron peak. Thus it makes sense to look at the abundances of some light elements in the outer layers,
that is in the layers that can be possibly ejected at the time of the supernova and see whether changes are seen depending on the rates used.

In Fig. \ref{fig:ejecta_content}, we compare the chemical composition 
in the supernova-ejecta of our different  models. We use the structure of the envelope obtained at the end of C-burning as a proxy for its structure at the pre-supernova stage.
The CO-core masses of those models (5.9 M$_\odot$) being almost equal (less than 0.1\% of difference), we deduced from the relation between CO-core and the remnant mass given in \citet{Maeder1992} a common remnant mass of 2.4 M$_\odot$. We compute the quantities of various isotopes present in all the layers above the remnant mass to obtain the numbers shown in Fig. \ref{fig:ejecta_content}.
%Figure \ref{fig:ejecta_content} shows the total abundances
%of a few elements in the supernova ejecta
%for our different 25\Msol\ models.
%{\bf M$_{\rm rem}$=2.38\Msol\ from M$_{\rm CO}$=5.87\Msol}. 
%\footnote{This value is deduced from the relation between the carbon-oxygen core mass, here 5.87\Msol, and the mass of the remnant given in \citet{Maeder1992}.}. 
Looking at the left panel, we see that the most affected elements are those which are the main products of C-burning, there are neon and magnesium. Their abundances are boosted in the Hin and HinRes models. Likely this is because the duration of the C-burning phase is slightly/significantly reduced in the HinRes/Hin models (see below) giving less time for these two elements to be destroyed by $\alpha$ captures (i.e. by the reactions $^{20}$Ne($\alpha$,$\gamma$)$^{24}$Mg($\alpha$,p)$^{27}$Al). 

%The Hin model produces much more $^{20}$Ne and $^{24}$Mg in the ejecta compared to the \citetalias{CF88} model and the HinRes model. However, in terms of mass content, it solely has an noticeable impact on $^{20}$Ne, linked to the $^{12}$C larger depletion. %In terms of mass content, the main difference is in the much larger $^{20}$Ne yield.

Table \ref{table:lifetimes} shows the central C-burning lifetimes, computed when 1\% of central $^{12}$C abundance has been burnt up to the central $^{12}$C abundance reaching value lower than 10$^{-5}$. The C-burning lifetimes are very similar between models with the \citetalias{CF88} rates and models using the HinRes reaction rates. Using solely the Hin reaction rates halves the lifetimes in comparison to the other models, due to the higher central temperature regime. 
Interestingly \citet{Chieffi2021}, using rates that are much higher than those of \citet{CF88} in the 0.1-1GK range,
find that C-burning occurs at lower densities \citep[see Fig. 3 in][]{Chieffi2021}. Indeed, since the nuclear rates are higher, a lower density is needed to reach the required amount of nuclear energy to counteract the contraction post He-b. This means that the beginning of C-burning is in fact happening earlier after the end of He-burning. This leads to a C-burning lifetime longer than the ones obtained with the CF88 rates. This behaviour is consistent with that seen in Fig.\ref{fig:rhoT}, where lower reaction rates led to reaching larger densities to burn Carbon, and hence a smaller C-b lifetime.

It is a well-known fact that the C-burning phase is the longest phase of the evolution of massive stars during which large neutrino emissions occurs \citep{Arnett1972,Woosley1986}. In Fig. \ref{fig:energy_production_25}, we can see that in the central regions of the 25\Msol\ model at the middle of the C-burning phase, the energy evacuated through neutrino emissions is nearly always superior to that produced by nuclear burning. Thus in that phase, the evolution of the stellar core
is driven mainly by neutrino emissions. The total quantity of entropy lost by the central region during the whole C-burning phase depends on the product of the time-averaged rate of neutrino emission during that phase and its duration. In the Hin model, the duration is shortened but on the other hand the 
entropy at the end of that phase not only depends on its duration but also on other factors such
as the neutrino emissions and the nuclear energy production rate. 
%In Fig. \ref{fig:energy_production_25} just a snapshot of the different type of energies (nuclear, neutrinos and gravitational) is shown at the middle of the carbon burning phase. 
%In the Hin model both the neutrino energy loss and the energy generation rates are lowered with respect to the other models. Making difficult to deduce which of the model lose the largest amount of entropy at that stage. W
In fact, we can see that the Hin model loses more energy through neutrino emission than the CF88 and HinRes models.   
We can moreover notice in Fig.~\ref{fig:rhoT}, that the Hin model crosses more rapidly the degeneracy limit than the other models, implying that it has lost more entropy than the other models during the C-burning phase and thus
becomes sensitive to degeneracy effects at earlier stages. This may have important consequences
for the ultimate fate of the star, the consequence of the core collapse and the nature of the stellar remnant \citep[see the interesting discussion in][where the link between the duration of the C-burning phase and these different aspects are discussed]{Woosley1986}.
We defer, however, a detailed discussion of this point to a future work where more advanced stages of the evolution will be computed using
a new version of GENEC accounting for the radial acceleration term and for a more detailed nuclear reaction network.

\section{Impact on detailed nucleosynthesis}

In this paragraph, we discuss the impact of the changes of the carbon fusion rates on the detailed composition resulting from C-burning.
For doing this, we use what we call here a ``one-layer model'' \citep{Choplin16}. 
Instead of computing the evolution of detailed stellar models composed 
of many hundreds/thousands of layers\footnote{A stellar model computes the evolution as a function of time of the physical quantities at different positions (layers) inside the star. In GENEC, the variables are temperature, pressure, luminosity, radius, chemical composition and the independent variable is the mass coordinate.}, we compute here the evolution of the abundances in only one layer. This simplification allows the use of a much more extended nuclear reaction network. The
nuclear reaction network in the ``one-layer model''  follows the evolution of the abundances of 1454 isotopes. The code receives as an input an initial distribution of abundances (i.e. the abundances at the beginning of the core carbon-burning phase) and an evolutionary path in the temperature/density plane representative of the core C-burning phase in a star. These input ingredients are taken from the CF88 25\Msol\ stellar model computed for the present paper. Note that using a 1454 elements network would be very computationally demanding if included in the build-up of stellar models. The stellar models are here computed with a reduced nuclear reaction network that accounts for all the reactions producing significant amounts of energy. Such reduced networks are sufficient to compute reliable stellar structures. On the other hand, this reduced network may miss some reactions that are not important for the energetic but have an impact on the abundances of some elements. This is what we want to check here by using a  ``one-layer model''. Three different ``one-layer models'' are computed. The first one uses the CF88 rates for the reactions $^{12}$C($^{12}$C,$\alpha$)$^{20}$Ne, $^{12}$C($^{12}$C,p)$^{23}$Na and $^{12}$C($^{12}$C,n)$^{23}$Mg. 
%The second model considers the hindrance phenomena (from this work) in these reactions. The third model the hindrance phenomena plus resonance.
The second and third models use rates from Hin model and HinRes model respectively for $^{12}$C($^{12}$C,$\alpha$)$^{20}$Ne and $^{12}$C($^{12}$C,p)$^{23}$Na, and we use the recent experimental rates from \citet{Bucher2015} for $^{12}$C($^{12}$C,n)$^{23}$Mg. 

\begin{figure}
     \centering
     \includegraphics[width=0.48\textwidth]{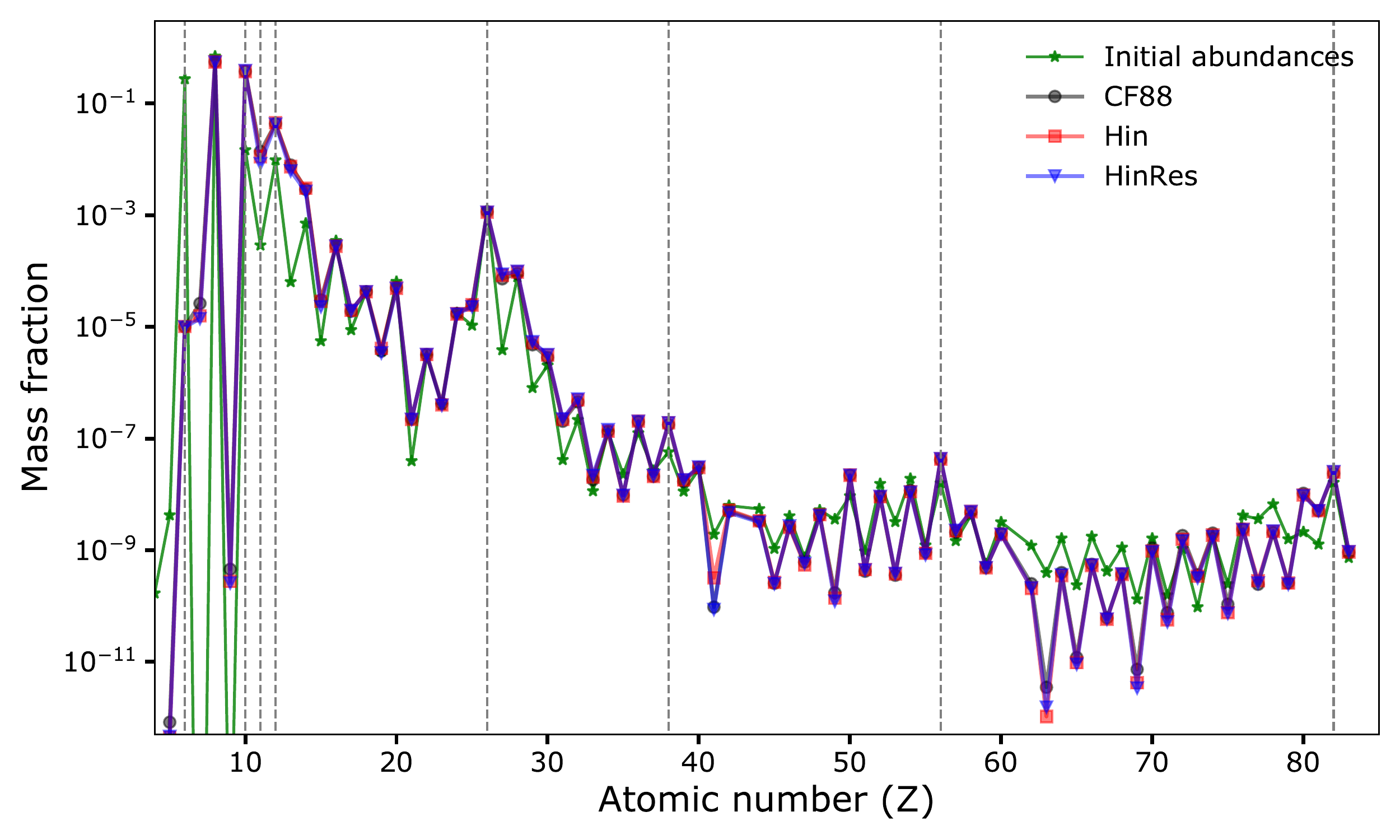}
     \caption{Comparisons between the abundances before (green) and at the end of the core C-burning phase (black, red and blue) obtained with three different sets of rates for the $^{12}$C~+~$^{12}$C reactions. The 8 dashed vertical lines highlight from left to right the carbon ($Z=6$), neon ($Z=10$), sodium ($Z=11$), magnesium ($Z=12$), iron ($Z=26$), strontium ($Z=38$), barium ($Z=56$) and lead ($Z=82$).}
     \label{box1}
 \end{figure}

Figure~\ref{box1} shows the mass fraction of the elements up to bismuth ($Z=83$) at the beginning of the C-burning phase (see the green pattern) and at the end of it using CF88, Hin, and HinRes rates (black, red and blue lines respectively). The comparison between the red and black/green/blue lines shows the impact of the C-burning. 
As expected we see that the two elements that are produced the most by carbon burning are $^{20}$Ne and $^{24}$Mg; $^{23}$Na is also produced
by this phase, but as indicated in the introduction, most of it has been transformed into $^{20}$Ne. Interestingly, we can see that strontium ($Z$ = 38)
is slightly enhanced during the C-burning phase. Strontium is produced by slow neutron capture, the neutrons being released by the reaction $^{13}$C($\alpha$, n)$^{16}$O reaction \citep{Pignatari2013}. Barium ($Z$ = 56) and lead, also produced by neutron capture are also slightly enriched.

\begin{figure}
     \centering
     \includegraphics[width=0.48\textwidth]{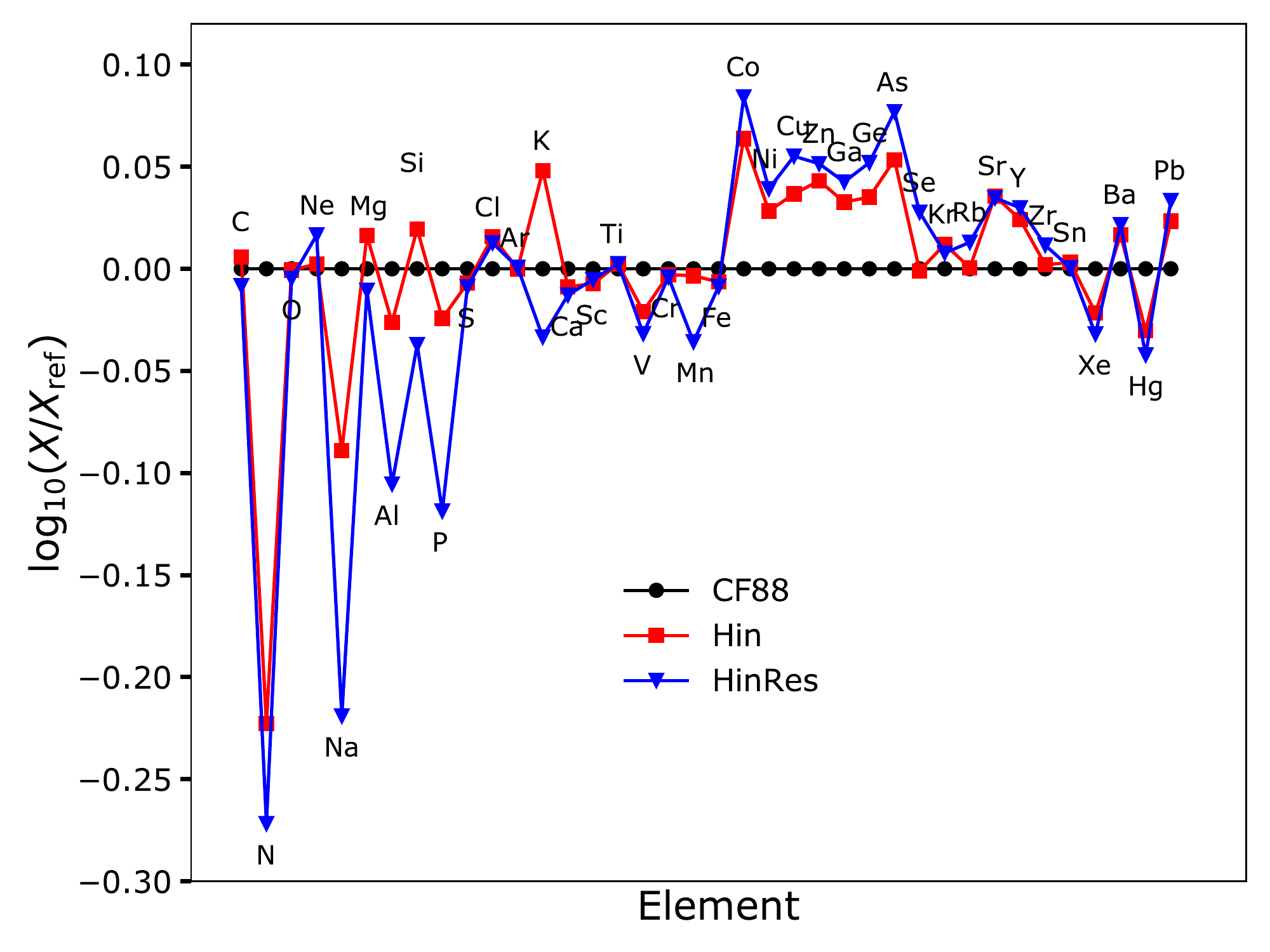}
     \caption{Abundances obtained at the end of the C-burning phase normalised to the final abundances obtained using the CF88 rate. Only the elements with a mass fraction greater than $10^{-8}$ (in either the first or second model) are considered.}
     \label{box2}
 \end{figure}

If we compare the final abundances obtained with the different rates for the $^{12}$C~+~$^{12}$C reaction, globally we see that the changes remain very modest and are hardly visible in Fig.~\ref{box1}. 
Figure~\ref{box2} represents the abundances normalised to the final abundances obtained using the CF88 rate. Only the elements with a significant mass fraction ($>10^{-8}$) are shown.

Very generally, we see that the final abundances obtained with the Hin rate show less difference with those obtained with the \citetalias{CF88} rate 
%for the low $Z$ ($<$ 20) elements 
than the abundances obtained with the HinRes rate. 
%On the other hand, the differences are more marked for higher $Z$ elements. 
The differences between the abundances obtained with the different rates never
go beyond a factor of 2. %60\%. 
After phosphorus (P), the biggest differences are about 20\% (for Co and As).
%The differences at high $Z$ (larger than 18) are less than 13\%.

To give a more detailed account, we first note that, as expected, since we show the situation when the carbon has been burnt, there is, by construction, no differences for the carbon abundances. Other abundant $\alpha$-nuclei like $^{16}$O, $^{20}$Ne, $^{24}$Mg are modestly affected in this ``one-layer'' model. 
The largest differences arise for nitrogen and sodium. The nitrogen mass fraction is however very small ($\sim 10^{-5}$) while the sodium mass fraction is $\sim 0.1$. Aluminium and phosphorus also show some differences. These differences are more remarkable for aluminium which has a mass fraction of about 0.01 (the phosphorus mass fraction is only about $10^{-5}$).
%The largest differences arise for odd elements such as $^{23}$Na ($Z$=11), $^{27}$Al ($Z$=13) and $^{31}$P ($Z$=15). 
%The nitrogen ($Z$=7) is also significantly affected but this  element has a small abundance (about $10^{-5}$). 
On the whole, these ``one-layer'' computations demonstrate 
%that for fixed temperature/density conditions, 
that the abundant element that is the most affected by a change of the carbon fusion rate is $^{23}$Na.
For the high $Z$ elements, the different rates 
do not affect too much their abundances at the end of the C-burning phase.

The abundance differences obtained in Fig.~\ref{box2} are too small to have an impact on the evolution and structure of the star during the last evolutionary stages. The nucleosynthesis may nevertheless be impacted: a different Na abundance after the end of core C-burning should impact the production of $^{26}$Mg through $^{23}$Na($\alpha$,$p$)$^{26}$Mg \citep{Arnett1974,Whitmire1974}. The protons given by this reaction should also increase the production of $^{26}$Al through $^{25}$Mg($p$,$\gamma$)$^{26}$Al \citep{Iliadis2011}.

When we look at the lifetime of such ``one-layer model'', we obtain a quite similar C-burning lifetime between the CF88 and the HinRes model (respectively 0.31 and 0.37 kyrs), due to the similar reaction rates in this temperature range. 
The Hin model lifetime is about 8 times longer than the CF88 model. This is due to the absence of the resonance in the Hin model. This lowers the $^{12}$C + $^{12}$C rate by about 1 dex at $0.8$~GK  (Fig.~\ref{fig:rr}). Consequently, the carbon takes more time to burn.
This is the opposite of what is obtained in stellar models, where the use of the Hin rate produces a shortening of the C-burning lifetime. 
This comes from the fact that, in the ``one-layer model'', there is no possibility for any feedback between the energy produced by the nuclear reactions and the path followed in the temperature/density plane. The path remains the same whatever rate is used.
%Indeed, the evolution in the temperature/density plane is predefined and the energy released by the nuclear reactions cannot change this path. 
Thus, the model with the lower rate simply takes more time to consume the carbon.

%The key explanation here comes from the neutrino energy loss. If we look at the stellar models, the lower rates of the Hind model leads to higher temperatures, leading then to a larger $\nu$-cooling. The core need now to produce a larger amount of energy to counterbalance both the gravitational collapse and the larger $\nu$-cooling. Hence, the carbon is burned faster than in the CF88 model. However in the box models, the condition of the layer are imposed to follow the same central conditions (i.e. the same temperatures through the whole C-burning phase). This means that the $\nu$-cooling is the same for each models, and that only the nuclear rates differ. The hin model having a much lower rate, it takes it much longer to consume the carbon than the CF88 and HinRes models, hence resulting in a very different outcome that the stellar models. 

%\begin{figure}
%     \centering
%     \includegraphics[width=0.5\textwidth]{Figure_3.pdf}
%     \caption{Same as Fig.~\ref{box2} but zoomed on the elements %with $Z < 21$.}
%     \label{box3}
% \end{figure}

\section{Summary and conclusions}

We have presented new $^{12}$C~+~$^{12}$C nuclear reaction rates both in the form of a table and of analytical fits, as well as uncertainties from the cross-section measurements. Experimental data were analyzed with Hin and HinRes response functions; the latter giving a more accurate approximation. Taking into account the width of the Gamow peak yields a sensitivity of the present experimental data to reaction rates down to temperatures of 0.60~GK, within one-sigma confidence intervals. In this temperature range, the interpolation of the cross sections measured by STELLA is the only input required to determine the reaction rates. 
%For lower temperatures, an extrapolation is needed.
This coincides with the effective C-burning temperatures region for both nuclear fusion scenarios investigated (i.e. Hin and HinRes scenarios), for stellar-evolution models of 12 and 25\Msol\ stars studied in this work. We therefore conclude that the direct measurement from \cite{Fruet2020} can provide the experimental input required for stellar evolution models. 
%We therefore conclude that the direct measurement from \cite{Fruet2020} {\bf is able to provide} the necessary sensitivity suited as an experimental input into stellar evolution models. 
%, {\bf such as} the 12 and 25\Msol\ models.
%that are responsible for ..\% of the cosmic inventory of element abundances.

Before we recall the main impacts of the new rates on the stellar models, let us remind here that the evolution of a star is mainly governed by gravity that tends to contract more and more the central regions. The contraction is stopped during the main nuclear phases because the nuclear reaction allows the pressure gradients to be maintained for long times. During the nuclear burning phases, the nuclear energy released per unit time and mass cannot change much. Indeed, any decrease of it implies a contraction, thus an increase of the temperature (in case of a perfect gas equation of state) and thus an increase of the energy released by nuclear reactions. Vice-versa, an excess of energy produced by nuclear reactions produces an expansion, a cooling and thus a decrease of the nuclear reaction rates. So at order zero, when a nuclear reaction rate important for the production of energy is changed, the energy released remains constant while the temperature changes. Of course, these temperature changes have repercussions on the detailed structure of the star and numerical models are needed to deduce the consequences. Note, however, that because nuclear reaction rates are very sensitive to very small change of the temperature, only a small change of the temperature needs to occur to compensate for a given change of the rate. This line of reasoning allows one to understand why even significant change in a nuclear reaction rate may have only modest effect on the stellar structure.

Although some of the rates differ by more than an order of magnitude at temperatures around 0.8 GK, the models outputs are only slightly changed. We find that in models computed with the lower rate (Hin case), the core C-burning occurs at central temperatures that are higher by about 10\%. As a consequence, the carbon burning phase is shortened by about a factor of two. 
This may impact the ultimate phases before the core collapse and have consequence for the type of event subsequent to that collapse (supernova or not) and for the nature of the stellar remnant (neutron star or black hole). We defer to a future work,
when our stellar evolution code will be adapted to the computation of these last stellar evolution phases, a detailed discussion on how these rates will impact the initial condition of the core collapse.
Concerning the composition of the supernova ejecta (assuming a neutron star is formed), the use of the lower Hin rate may impact some isotopes such as $^{20}$Ne that are found to be enhanced by an order of magnitude in models computed with it.

In addition to the investigations carried out on the impact of the reaction rates determined in this work on the C-burning phase of massive stars, other studies of their impacts could be made on cases where the carbon fusion reaction plays a key role at lower temperatures, where the discrepancies with older rates is the greatest, such as the C-burning in intermediate-mass stars, in type Ia supernovae, or even in superbursts \citep{Strohmayer2002}.

\begin{acknowledgements}
JN acknowledges support from the Interdisciplinary Thematic Institute QMat, as part of the ITI 2021 2028 program of the University of Strasbourg, CNRS and Inserm, which was supported by IdEx Unistra (ANR 10 IDEX 0002), and by SFRI STRAT’US project (ANR 20 SFRI 0012) and EUR QMAT ANR-17-EURE-0024 under the framework of the French Investments for the Future Program. SM, GM, and SE acknowledge the STAREX grant from the ERC Horizon 2020 research and innovation programme (grant agreement No 833925), and the COST Action ChETEC (CA 16117) supported by COST (European Cooperation in Science and Technology). This work was supported by the Fonds de la Recherche Scientifique-FNRS under Grant No IISN 4.4502.19.
\end{acknowledgements}

\bibliographystyle{aa}
\bibliography{rr}{}

\appendix
\section{Tables and analytic formulae}

Table \ref{tab:rr} summarizes values of reaction rates for different temperatures. Recommended rates are based on a classical determination of the reaction rates using cross-section extrapolation, and lower and upper rates represent the lower and upper limits of recommended rates, using their 1$\sigma$ uncertainties. The latter comes from experimental uncertainties on cross sections measured by STELLA experiment \citep{Fruet2020}.

\begin{table*}
\centering
\caption{Reaction rates for the $^{12}$C~+~$^{12}$C fusion reaction according to different models. Reaction rates are in $ \mathrm{cm} ^{3} \mathrm{s}^{-1} \mathrm{mol}^{-1} $ and temperatures in GK.}
\label{tab:rr}
\begin{tabular}{lcccccc}
    \hline \hline
    \noalign{\smallskip}
    T & \multicolumn{3}{c}{Hin model} & \multicolumn{3}{c}{HinRes model} \\
    \noalign{\smallskip}
    \cline{2-4}
    \cline{5-7}
    \noalign{\smallskip}
      & Recomm. & Lower & Upper & Recomm. & Lower & Upper \\
    \noalign{\smallskip}
    \hline
    \noalign{\smallskip}
        0.11 & 1.79E-54 & 1.15E-54 & 2.42E-54 & 2.94E-54 & 1.87E-54 & 4.00E-54 \\
        0.12 & 4.40E-52 & 2.89E-52 & 5.90E-52 & 7.02E-52 & 4.58E-52 & 9.46E-52 \\
        0.13 & 6.03E-50 & 4.05E-50 & 8.02E-50 & 9.39E-50 & 6.26E-50 & 1.25E-49 \\
        0.14 & 5.10E-48 & 3.48E-48 & 6.71E-48 & 7.75E-48 & 5.26E-48 & 1.02E-47 \\
        0.15 & 2.87E-46 & 1.99E-46 & 3.74E-46 & 4.27E-46 & 2.94E-46 & 5.60E-46 \\
        0.16 & 1.14E-44 & 8.03E-45 & 1.48E-44 & 1.67E-44 & 1.17E-44 & 2.17E-44 \\
        0.18 & 7.68E-42 & 5.54E-42 & 9.82E-42 & 1.08E-41 & 7.78E-42 & 1.39E-41 \\
        0.2 & 2.08E-39 & 1.53E-39 & 2.63E-39 & 2.86E-39 & 2.10E-39 & 3.62E-39 \\
        0.25 & 1.53E-34 & 1.17E-34 & 1.89E-34 & 1.98E-34 & 1.51E-34 & 2.45E-34 \\
        0.3 & 7.67E-31 & 6.05E-31 & 9.29E-31 & 9.54E-31 & 7.50E-31 & 1.16E-30 \\
        0.35 & 6.77E-28 & 5.45E-28 & 8.08E-28 & 8.16E-28 & 6.56E-28 & 9.77E-28 \\
        0.4 & 1.79E-25 & 1.47E-25 & 2.11E-25 & 2.16E-25 & 1.76E-25 & 2.57E-25 \\
        0.45 & 1.97E-23 & 1.64E-23 & 2.30E-23 & 2.77E-23 & 2.20E-23 & 3.34E-23 \\
        0.5 & 1.11E-21 & 9.37E-22 & 1.29E-21 & 2.34E-21 & 1.75E-21 & 2.92E-21 \\
        0.55 & 3.75E-20 & 3.18E-20 & 4.31E-20 & 1.27E-19 & 9.08E-20 & 1.63E-19 \\
        0.6 & 8.34E-19 & 7.15E-19 & 9.54E-19 & 4.13E-18 & 2.88E-18 & 5.38E-18 \\
        0.65 & 1.32E-17 & 1.14E-17 & 1.51E-17 & 8.30E-17 & 5.75E-17 & 1.09E-16 \\
        0.7 & 1.59E-16 & 1.38E-16 & 1.80E-16 & 1.11E-15 & 7.67E-16 & 1.45E-15 \\
        0.75 & 1.51E-15 & 1.32E-15 & 1.70E-15 & 1.06E-14 & 7.35E-15 & 1.39E-14 \\
        0.8 & 1.18E-14 & 1.03E-14 & 1.32E-14 & 7.70E-14 & 5.36E-14 & 1.00E-13 \\
        0.85 & 7.73E-14 & 6.81E-14 & 8.64E-14 & 4.46E-13 & 3.14E-13 & 5.79E-13 \\
        0.9 & 4.37E-13 & 3.87E-13 & 4.87E-13 & 2.15E-12 & 1.53E-12 & 2.77E-12 \\
        0.95 & 2.17E-12 & 1.93E-12 & 2.41E-12 & 8.93E-12 & 6.44E-12 & 1.14E-11 \\
        1 & 9.62E-12 & 8.57E-12 & 1.07E-11 & 3.28E-11 & 2.41E-11 & 4.15E-11 \\
        1.05 & 3.85E-11 & 3.44E-11 & 4.27E-11 & 1.09E-10 & 8.15E-11 & 1.36E-10 \\
        1.1 & 1.41E-10 & 1.26E-10 & 1.56E-10 & 3.34E-10 & 2.55E-10 & 4.12E-10 \\
        1.15 & 4.77E-10 & 4.27E-10 & 5.26E-10 & 9.58E-10 & 7.51E-10 & 1.16E-09 \\
        1.2 & 1.50E-09 & 1.34E-09 & 1.65E-09 & 2.61E-09 & 2.10E-09 & 3.13E-09 \\
        1.25 & 4.40E-09 & 3.96E-09 & 4.84E-09 & 6.81E-09 & 5.59E-09 & 8.03E-09 \\
        1.3 & 1.22E-08 & 1.10E-08 & 1.34E-08 & 1.71E-08 & 1.43E-08 & 1.99E-08 \\
        1.35 & 3.20E-08 & 2.88E-08 & 3.51E-08 & 4.14E-08 & 3.53E-08 & 4.76E-08 \\
        1.4 & 7.99E-08 & 7.21E-08 & 8.76E-08 & 9.72E-08 & 8.39E-08 & 1.11E-07 \\
        1.45 & 1.90E-07 & 1.72E-07 & 2.09E-07 & 2.21E-07 & 1.93E-07 & 2.49E-07 \\
        1.5 & 4.36E-07 & 3.94E-07 & 4.77E-07 & 4.87E-07 & 4.29E-07 & 5.45E-07 \\
        1.75 & 1.61E-05 & 1.46E-05 & 1.76E-05 & 1.65E-05 & 1.49E-05 & 1.82E-05 \\
        2 & 3.01E-04 & 2.72E-04 & 3.29E-04 & 3.02E-04 & 2.73E-04 & 3.31E-04 \\
        2.5 & 2.70E-02 & 2.44E-02 & 2.96E-02 & 2.70E-02 & 2.43E-02 & 2.96E-02 \\
        3 & 7.45E-01 & 6.69E-01 & 8.22E-01 & 7.48E-01 & 6.71E-01 & 8.25E-01 \\
        3.5 & 9.69E+00 & 8.63E+00 & 1.08E+01 & 9.79E+00 & 8.71E+00 & 1.09E+01 \\
        4 & 7.52E+01 & 6.65E+01 & 8.39E+01 & 7.65E+01 & 6.75E+01 & 8.54E+01 \\
        5 & 1.63E+03 & 1.42E+03 & 1.84E+03 & 1.68E+03 & 1.46E+03 & 1.90E+03 \\
        6 & 1.47E+04 & 1.26E+04 & 1.68E+04 & 1.54E+04 & 1.32E+04 & 1.77E+04 \\  
        \noalign{\smallskip}
        \hline
\end{tabular}
\end{table*}

Reaction rates have been fitted with the formula from \citet{Thielemann}, used in the REACLIB library:
\begin{eqnarray}
    N_{A} \langle \sigma v \rangle & = & \sum_{i} \exp(a_{i0} + a_{i1}/T_{9} + a_{i2}/T_{9}^{1/3} + a_{i3} T_{9}^{1/3} \nonumber \\ 
    & & + a_{i4} T_{9} + a_{i5} T_{9}^{5/3} + a_{i6} ln(T_{9})),
\end{eqnarray}
where $a_{i}$ are fitting parameters and $T_{9}$ is the temperature in GK.
In the case of the Hin model, the best fit was obtained with $i=1$, and for HinRes model with $i=2$. The different parameters are reported in Tables \ref{tab:par_rr_hin} and \ref{tab:par_rr_hin_res} for the Hin model and the HinRes model respectively.

\begin{table}
    \centering
    \caption{Parameters for recommended, lower and upper reaction rates in the case of Hin model.}
    \label{tab:par_rr_hin}
    \begin{tabular}{lccc}
        \hline \hline
        \noalign{\smallskip}
        $a_{ij}$ & Recomm. & Lower & Upper \\
        \noalign{\smallskip}
        \hline
        \noalign{\smallskip}
        $a_{10}$ & 7.72E+01 & 7.77E+01 & 7.68E+01 \\
        $a_{11}$ & 6.18E-02 & 6.39E-02 & 4.41E-02 \\
        $a_{12}$ & -9.45E+01 & -9.50E+01 & -9.40E+01 \\
        $a_{13}$ & -7.73E+00 & -7.83E+00 & -7.64E+00 \\
        $a_{14}$ & -4.35E-01 & -4.57E-01 & -4.16E-01 \\
        $a_{15}$ & 3.11E-02 & 3.33E-02 & 2.88E-02 \\
        $a_{16}$ & 2.28E-01 & 1.64E-01 & 2.88E-01 \\
        \noalign{\smallskip}
        \hline
    \end{tabular}
\end{table}

\begin{table}
    \centering
    \caption[]{Parameters for recommended, lower and upper reaction rates in the case of HinRes model.}
    \label{tab:par_rr_hin_res}
    \begin{tabular}{lccc}
        \hline \hline
        \noalign{\smallskip}
        $a_{ij}$ & Recomm. & Lower & Upper \\
        \noalign{\smallskip}
        \hline
        \noalign{\smallskip}
        $a_{10}$ & 7.63E+01 & 7.68E+01 & 7.58E+01 \\
        $a_{11}$ & 4.97E-02 & 5.12E-02 & 3.18E-02 \\
        $a_{12}$ & -9.37E+01 & -9.42E+01 & -9.32E+01 \\
        $a_{13}$ & -7.60E+00 & -7.70E+00 & -7.51E+00 \\
        $a_{14}$ & -4.19E-01 & -4.44E-01 & -3.97E-01 \\
        $a_{15}$ & 2.97E-02 & 3.21E-02 & 2.72E-02 \\
        $a_{16}$ & 3.53E-01 & 2.95E-01 & 4.10E-01 \\
        $a_{20}$ & 3.42E-01 & 5.07E-01 & -3.62E+00 \\
        $a_{21}$ & -2.48E+01 & -2.48E+01 & -2.52E+01 \\
        $a_{22}$ & 3.25E-01 & 3.97E-01 & 3.94E+00 \\
        $a_{23}$ & -3.65E-01 & -3.32E-01 & 5.97E-02 \\
        $a_{24}$ & 1.64E-02 & -7.59E-03 & -1.45E-01 \\
        $a_{25}$ & -7.65E-04 & 1.68E-03 & 1.23E-02 \\
        $a_{26}$ & -1.29E+00 & -1.24E+00 & -4.16E-01 \\
        \noalign{\smallskip}
        \hline
    \end{tabular}
\end{table}

%\end{appendix}

\end{document}